\documentclass[%
reprint,amsmath,amssymb,aps,prl,superscriptaddress,
twocolumn
]{revtex4-1}

\usepackage{graphicx}
\usepackage{float}
\usepackage{subcaption}
\begin{document}
	
	\title{Electrical generation and detection of spin waves in a quantum Hall ferromagnet}
	
	\author{Di S. Wei}
	\affiliation{John A. Paulson School of Engineering and Applied Sciences, Harvard University, Cambridge, Massachusetts 02138}
	
	\author{Toeno van der Sar}
	\affiliation{Department of Physics, Harvard University, Cambridge, Massachusetts 02138}
	
	\author{Seung Hwan Lee}
	\affiliation{Department of Physics, Harvard University, Cambridge, Massachusetts 02138}
		
	\author{Kenji Watanabe}
	\affiliation{Advanced Materials Laboratory, National Institute for Materials Science, Tsukuba, Ibaraki 305-0044, Japan}
	
	\author{Takashi Taniguchi}
	\affiliation{Advanced Materials Laboratory, National Institute for Materials Science, Tsukuba, Ibaraki 305-0044, Japan}
		
	\author{Bertrand I. Halperin}
	\affiliation{Department of Physics, Harvard University, Cambridge, Massachusetts 02138}
		
	\author{Amir Yacoby}
	\affiliation{John A. Paulson School of Engineering and Applied Sciences, Harvard University, Cambridge, Massachusetts 02138}
	\affiliation{Department of Physics, Harvard University, Cambridge, Massachusetts 02138}

	 \begin{abstract} 
	 	
	 	Spin waves are collective excitations of magnetic systems. An attractive setting for studying long-lived spin-wave physics is the quantum Hall (QH) ferromagnet, which forms spontaneously in clean two-dimensional electron systems at low temperature and in a perpendicular magnetic field. We used out-of-equilibrium occupation of QH edge channels in graphene to excite and detect spin waves in magnetically ordered QH states. Our experiments provide direct evidence for long distance spin wave propagation through different ferromagnetic phases in the N=0 Landau level, as well as across the insulating canted antiferromagnetic phase. Our results will enable experimental investigation of the fundamental magnetic properties of these exotic two-dimensional electron systems.
	 \end{abstract}

	  	\maketitle
	  	
	Quantum Hall (QH) ferromagnetism arises from the interaction of electrons in massively degenerate, quantized energy levels known as Landau levels (LLs) \cite{Girvin2000}. When disorder is low enough for Coulomb interactions to manifest, the electrons in partially filled LLs spin-polarize spontaneously to minimize their exchange energy, with the single-particle Zeeman effect dictating their polarization axis \cite{Young2012, Sondhi1993}. In graphene, these phenomena give rise to ferromagnetic phases when the N=0 LL is at quarter- and three-quarter-filling \cite{Alicea2006, Yang2006, Zhang2006, Nomura2008, Goerbig2011}. Such QH ferromagnets have an insulating topological bulk and spin-polarized edge states. Furthermore, a canted antiferromagnetic (CAF) state is believed to emerge at one-half filling, with a canting angle determined by the competing valley anisotropy and Zeeman energy \cite{Kharitonov2012, Young2014}. Spin waves, also known as magnons, are the lowest energy excitation in both the QH ferromagnet and CAF \cite{Girvin2000, Green2002, Takei2016}, and could provide crucial information about these topologically non trivial magnetic states. 
	
	In our experimental setup, we generate magnons by creating an imbalance of chemical potential between two edge states of opposite spin that run along the boundary of a QH magnet. If this imbalance is smaller than the energy required for generating magnons in the QH magnet (and there are no thermal magnons already present in the system), scattering between these two edge states is forbidden because the change in angular momentum of a scattered electron cannot be absorbed by the system. Indeed, previous measurements have shown that oppositely spin-polarized edge channels do not equilibrate as long as the imbalance is small \cite{Amet2014, Wei2017}. However, we find edge channel equilibration commences when the imbalance exceeds the minimum energy required for exciting magnons in the QH ferromagnet. Because the magnetization of the QH ferromagnet is extremely dilute, there are negligible demagnetizing fields and the minimum energy to excite magnons is given by the Zeeman energy $E_\mathrm{Z}=g\mu_B B$ \cite{Girvin2000, Kittel1948}, where $g$ is the electron g-factor, $\mu_B$ is the Bohr magneton, and $B$ is the external magnetic field. Although magnon generation does not directly affect the conductance of the system, the reverse process of magnon absorption by far-away edge states does, allowing us to detect the propagation of magnons electrically, in close analogy to the conventional detection of magnons in insulators via the inverse spin Hall effect \cite{Kajiwara2010, Cornelissen2015, Chumak2015, Wesenberg2017}.
	
	\begin{figure*}
		\includegraphics[width=0.75\textwidth]{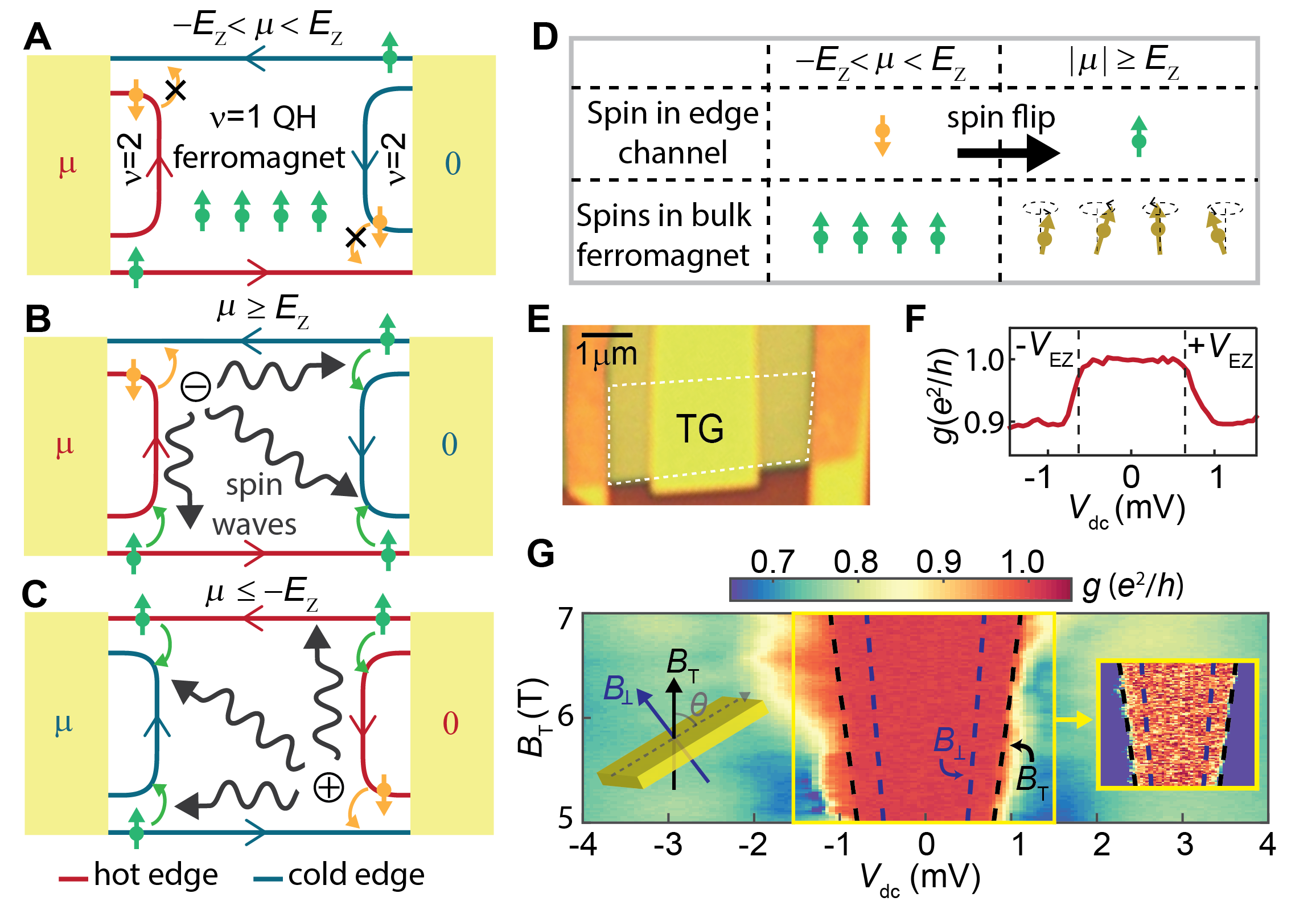}%
		\caption{\textbf{Magnons in a quantum Hall ferromagnet.} \textbf{(A-C)} A chemical potential difference ($\mu$) is applied between the left and right leads. Edge channels with high and low chemical potential are labeled  ``hot'' and ``cold'', respectively. Spin-up and spin-down polarization is denoted by the green and orange arrows, respectively. The central region is tuned to $\nu$ = 1 and adjacent regions are tuned to $\nu$ = 2. \textbf{(A)} The chemical potential difference ($\mu$) between the spin-up and spin-down edge channel is less than the Zeeman energy ($E_\mathrm{Z}$), and scattering is suppressed. \textbf{(B)} $\mu \geq E_\mathrm{Z}$: Electrons have enough energy to flip their spins and transfer spin angular momentum (magnons) into the bulk (at the encircled minus sign). These magnons are absorbed at distant corners, causing electrons to flip from spin-up into spin-down channels. \textbf{(C)} $\mu \leq -E_\mathrm{Z}$: Magnons are generated at the location denoted by the encircled plus sign. \textbf{(D)} Bulk spin polarization before and after magnon creation, conserving total spin angular momentum. \textbf{(E)} Optical micrograph of device 1; graphene is outlined in white. TG, top gate. \textbf{(F)} A dc voltage ($V_\mathrm{dc}$) and a 50-$\mu$V ac excitation voltage ($V_\mathrm{ac}$) are applied to the left contact and the differential conductance (d$I$/d$V$, where $V = V_\mathrm{ac}+V_\mathrm{dc}$) is measured through the right contact ($B_\mathrm{perp}$ = 4 T, $V_{TG}$ = -0.18 V, $V_{BG}$ = 3 V).  Conductance is quantized to $e^2/h$ until $|\mu| \geq ≥ E_\mathrm{Z}$. \textbf{(G)} d$I$/d$V$ as a function of bias and magnetic field. The blue dashed line is the Zeeman energy, $E_\mathrm{Z\perp} = g\mu_B B_\perp$ calculated using the perpendicular (total) magnetic field $B_\perp$; The black dashed line is the Zeeman energy, $E_\mathrm{ZT}=g \mu_\mathrm{B} B_\mathrm{T}$ calculated using the total magnetic field $B_\mathrm{T}$). Both the top gate ($V_{TG}$) and back gate ($V_{BG}$) are swept to stay at $nu=1$ throughout the device from 7T ($V_{TG}$ = 0.16 V, $V_{BG}$ = 0.73 V) to 5T ($V_{TG}$ = 0.12 V, $V_{BG}$ = 0.44 V)  The decrease in conductance from $e^2/h$ evolves linearly with the magnetic field coinciding with $E_\mathrm{ZT}$ rather than $E_{\mathrm{Z}\perp}$. Right inset: A saturated color plot (from 0.98 to 1.02 $e^2/h$) of the region enclosed by the yellow box. All measurement are conducted in a cryostat with a base temperature of 20 mK.}
		\label{fig:F3}
	\end{figure*} 

	To demonstrate spin wave propagation, we begin with a dual-gated monolayer graphene device (device $1$) where the central region can be tuned to a different filling factor than the adjacent regions (Fig.\ 1A). Connecting the two leads is a chiral edge state that carries spin-polarized electrons aligned with the magnetic field, which we call spin-up. We tune the central region to a three-quarters-filled LL ($\nu = 1$), whereas the outer regions are tuned to a non-magnetic fully filled LL ($\nu = 2$). We apply a source-drain voltage $V_\mathrm{dc}$ to induce a difference in chemical potential $\mu  = -e V_\mathrm{dc}$ between the edge channels emerging from the two contacts, where $e$ is the electron charge.  Once $|\mu| \geq E_\mathrm{Z}$, an electron traveling in a high-energy (``hot''), spin-down edge state can relax into a low-energy (``cold''), spin-up edge state by emitting a magnon into the ferromagnetic bulk (Fig.\ 1, B and C). Because equilibration must occur close to the ferromagnetic bulk in order to launch magnons, the edge states must equilibrate over short length scales at localized ``hot spots'' where the hot and cold edges meet. This makes graphene an ideal platform to observe this phenomenon, where edge state equilibration can occur over length scales $< 1 \mu$m \cite{Amet2014, Williams2007, Ozyilmaz2007} (See \cite {WeiSupp2018} for further discussion). Because only spin-down angular momentum can be propagated into the spin-up bulk, magnon generation occurs at the location denoted by an encircled minus sign when $\mu \geq E_\mathrm{Z}$ (Fig.\ 1B) and at the location denoted by an encircled plus sign when $\mu \leq -E_\mathrm{Z}$ (Fig.\ 1C).  These magnons propagate through the insulating QH ferromagnet and can be absorbed by the reverse process between other edge channels (Fig.\ 1B-C), which causes a deviation in the conductance from a well-quantized $\nu = 1$ QH state.

	When we measure the conductance of the graphene device (Fig.\ 1E, atomic force microscopy image in fig.\ S3) as a function of $V_\mathrm{dc}$, we find that the $\nu = 1$ QH ferromagnet remains precisely quantized at the expected value of $e^2/h$, and then changes once the applied bias reaches the Zeeman threshold ($V_\mathrm{dc}= \pm V_\mathrm{EZ}= \mp E_\mathrm{Z}/e$), as expected from our model (Fig.\ 1F). Interestingly, we find that thanks to contact doping \cite{WeiSupp2018, Giovannetti2008} we can tune the entire device to $\nu = 1$ and find the same phenomenon of conductance deviation at the Zeeman threshold (fig.\ S4).

	\begin{figure*}
		\includegraphics[width=1\textwidth]{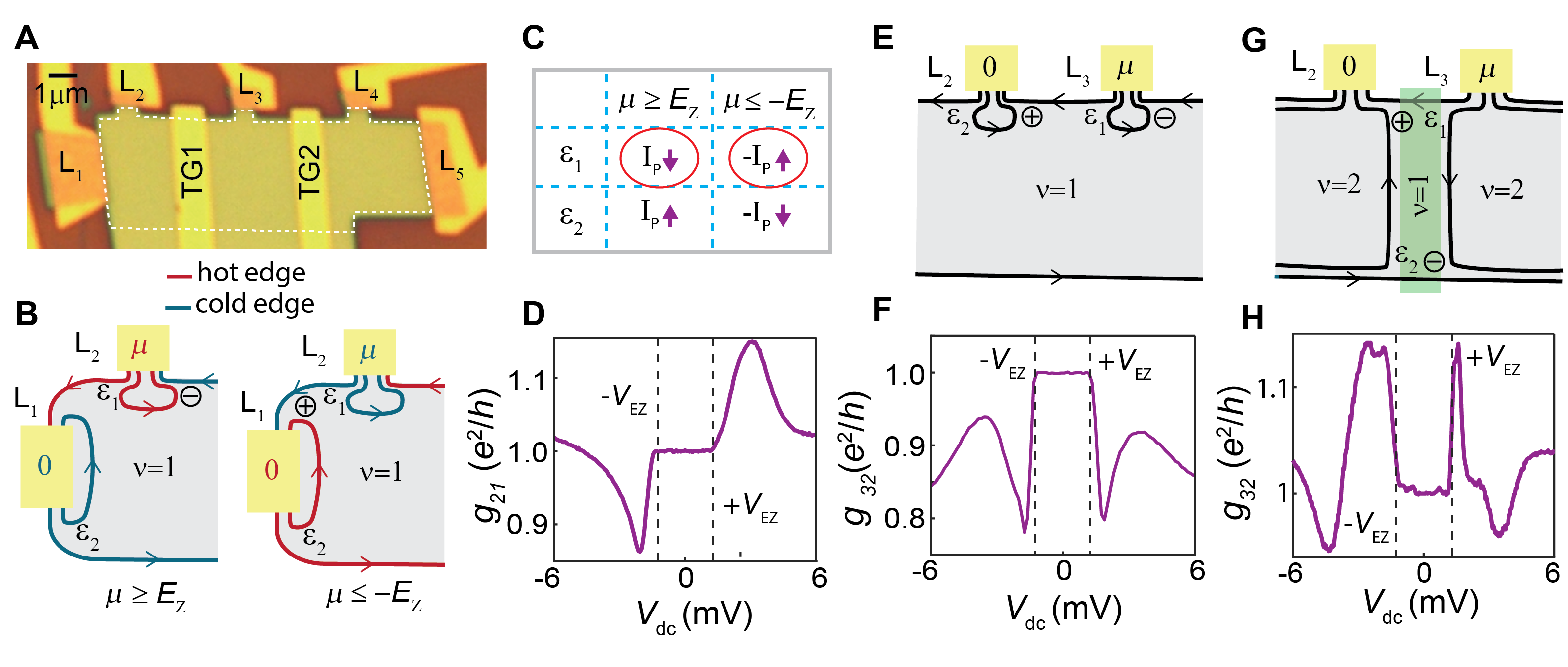}%
		\caption{\textbf{Effect of relative magnon absorption on conductance.} \textbf{(A)} Optical micrograph of device 2. Graphene is outlined in white. \textbf{(B)} Schematic of a two-terminal conductance measurement using leads L$_2$ and L$_1$ where hot and cold edges are colored red and blue, respectively, for both $\mu \geq E_\mathrm{Z}$ (left) and $\mu \leq -E_\mathrm{Z}$ (right), and the magnon generation site is labeled by the encircled plus or minus sign indicating positive or negative bias.  $\mu \geq E_\mathrm{Z}$ : magnon absorption at $\varepsilon_1$ transfers chemical potential from a forward-moving edge to a backward-moving edge, causing the particle current ($I_\mathrm{P} = -I/e$) to decrease. Conversely, magnon absorption at $\varepsilon_2$ transfers chemical potential from a backward-moving edge to a forward-moving edge, increasing $I_\mathrm{P}$. $\mu \leq -E_\mathrm{Z}$ : Magnon absorption at $\varepsilon_1$ causes an increase in $|-I_\mathrm{P}|$; absorption at $\varepsilon_2$ causes a decrease in $|-I_\mathrm{P}|$. \textbf{(C)} The effects of $\varepsilon_1$ and $\varepsilon_2$ at $\mu \geq E_\mathrm{Z}$ and $\mu \leq -E_\mathrm{Z}$. The current changes caused by $\varepsilon_1$ are dominant and are circled in red. The purple arrows indicate an increase (up) or decrease (down) in the magnitude of the signed particle current.\textbf{(D)} Conductance from $L_2$ to $L_1$ ($g_{21}$ = d$I$/d$V$ = d$I_\mathrm{P}$/d$\mu$) decreases at $V_\mathrm{dc} = -V_\mathrm{EZ}$ and increases at $V_\mathrm{dc} = V_\mathrm{EZ}$, indicating that $\varepsilon_1$ has a larger effect than $\varepsilon_2$ (B = 8 T, $V_{BG}$ = 4 V). See \cite{WeiSupp2018} for full circuit analysis. \textbf{(E-F)} Conductance from L$_3$ and L$_2$ ($g_{32}$) where the entire device is tuned to $\nu=1$ ($V_{BG}$ = 4V, TG1 = 0V is not shown). At positive bias, $\varepsilon_2 > \varepsilon_1$, and at negative bias, $\varepsilon_1 > \varepsilon_2$, resulting in a conductance drop for both biases.\textbf{(G-H)} Conductance from $L_3$ to $L_2$ ($g_{32}$) where TG1 is tuned to $\nu_\mathrm{TG1} = 1$ (TG1=-0.36 V) while the regions outside are set to $\nu_\mathrm{bg} = 2$ ($V_{BG}$ = 6.5V). At positive bias, $\varepsilon_1 > \varepsilon_2$, and at negative bias, $\varepsilon_2 > \varepsilon_1$, resulting in a conductance rise for both biases. See fig.\ S5. for a detailed analysis.}
		\label{fig:F2}
	\end{figure*} 

	\begin{figure*}
		\includegraphics[width=0.95\textwidth]{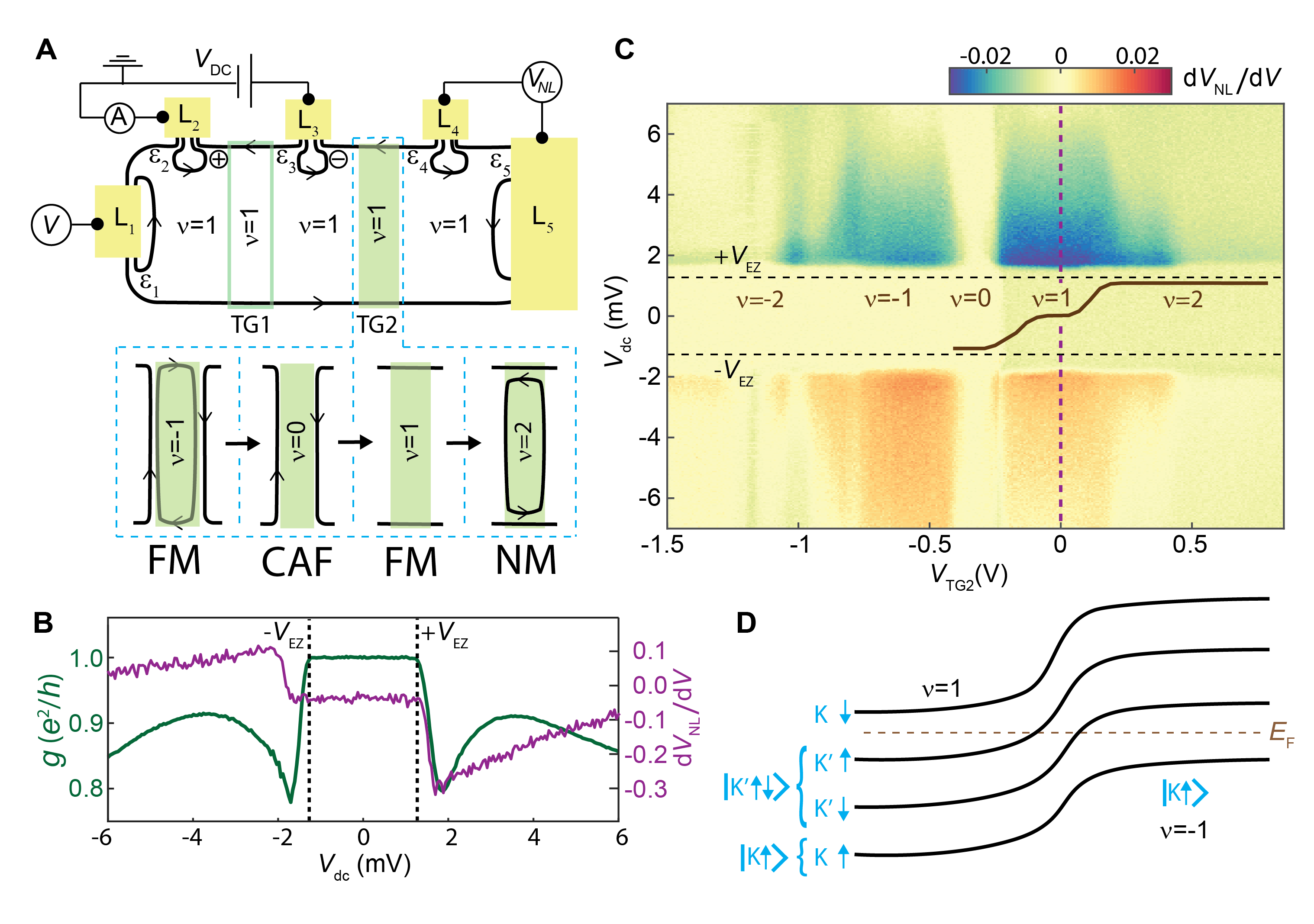}%
		\caption{\textbf{Non-local voltage signal due to magnon absorption.} Shown are the data from device 2. \textbf{(A)} Schematic circuit configuration for measuring a non-local voltage in device 2. The filling factor under TG1 ($\nu_\mathrm{TG1}$) = 1 for all measurements while the filling factor under TG2 ($\nu_\mathrm{TG2}$) is swept from -2 to 2, and the rest of the device is kept at $\nu_\mathrm{bg}$ = 1 ($V_{BG}$ = 4 V). The bottom panel highlights the magnetic properties of different cases of $\nu_\mathrm{TG2}$: non-magnetic (NM), ferromagnetic (FM), or canted antiferromagnetic (CAF). \textbf{(B)} $S_\mathrm{NL}$ (purple) superimposed onto $\mathrm{d}I/\mathrm{d}V$ (green) as a function of $V_\mathrm{dc}$ when $\nu_\mathrm{TG2}$ = 1 ($B$ = 8 T). The onset of $S_\mathrm{NL}$ is slightly offset in bias from the decrease in conductance, indicating that magnon generation needs to reach a threshold before being absorbed in distant contacts.  \textbf{(C)} A pronounced $S_\mathrm{NL}$ signal when $\nu_\mathrm{TG2}$ = 1 and $\nu_\mathrm{TG2}$ = -1 (See Fig. S8 for similar measurements using TG1). Tuning TG2 to the nonmagnetic QH phases ($\nu_\mathrm{TG2}$ = 2 and $\nu_\mathrm{TG2}$ = -2), as well the $\nu_\mathrm{TG2}$ = 0 CAF state, strongly suppresses $S_\mathrm{NL}$. There is a small finite background $S_\mathrm{NL}$ when edge states pass through TG2, discussed in fig.\ S7, B.  Solid brown line indicates where $\nu_\mathrm{TG2}$ = 0, 1, and 2 (fig.\ S7, C and D).  \textbf{(D)} The spatial variation of the LLs at a $\nu$ = 1/$\nu$ = -1 junction, with the expected valley and spin polarizations of each level labeled.}
		\label{fig:F3}
	\end{figure*}

By tilting the external magnetic field with respect to the sample-plane normal axis, we verify that the change in conductance occurs when the applied chemical potential exceeds the bare Zeeman energy $E_\mathrm{Z} = g \mu_\mathrm{B} B_\mathrm{T}$ ($g$=2), which is given by the total field $B_\mathrm{T}$ (Fig. 1G -- sample is tuned entirely to $\nu=1$). In contrast, previous transport studies of spin and valley excitations in graphene and GaAs have only found excitations related to the exchange energy gap \cite{Young2012, Sondhi1993, Schmeller1995}, which depends on the component of the field perpendicular to the sample plane ($B_\perp$). Our tilted-field measurements therefore corroborate our magnon-based interpretation of the observed change in sample conductance. All subsequent experiments described in this work are done at perpendicular field.

The conductance change at $E_\mathrm{Z}$ can either be positive or negative, depending on the number of magnons absorbed at each contact. To examine this, we use different sets of leads in the same device (Fig.\ 2A, device 2) to perform two-terminal conductance measurements. We start with leads L$_2$ and L$_1$ in Fig.\ 2B. We label the amount of redistributed chemical potential at each of the absorption sites $\varepsilon_{i}$, with $i$ indexing the absorption site (note that $\varepsilon_{i}$ = 0 for $-E_\mathrm{Z} < \mu < +E_\mathrm{Z}$), where $\varepsilon_{i}$ is proportional to the number of magnons absorbed at site $i$. Absorption at $\varepsilon_1$ and $\varepsilon_2$ have opposite effects on the conductance, as magnon absorption transfers chemical potential from the outer edge to the inner edge. Therefore, for $\mu \geq E_\mathrm{Z}$, magnon absorption at $\varepsilon_1$ decreases the particle current ($I_\mathrm{P} = - I/e$ where $I$ is the charge current) whereas magnon absorption at $\varepsilon_2$ increases $I_\mathrm{P}$ (Fig.\ 2B). For $\mu \leq -E_\mathrm{Z}$, the hot and cold reservoirs are reversed, and we now consider the change to the negative particle current $-I_\mathrm{P}$. Although $\varepsilon_1$ still decreases the particle current, $I_\mathrm{P}$ is now negative, and so $\varepsilon_1$ actually increases the magnitude of the particle current ($|-I_\mathrm{P}|$); similarly, for $\mu \leq -E_\mathrm{Z}$, $\varepsilon_2$ decreases $|-I_\mathrm{P}|$ (Fig.\ 2C). We can quantify this using current conservation to formulate the differential conductance as a function of $\varepsilon_i$ and $\mu$:

\begin{equation} \label{eq1}
\frac{\mathrm{d}I}{\mathrm{d}V} = \frac{\mathrm{d}I_\mathrm{P}}{\mathrm{d}\mu} 
= \frac{1}{R_\mathrm{Q}}   \Big( 1+\frac{\mathrm{d}\varepsilon_2}{\mathrm{d}\mu}-\frac{\mathrm{d}\varepsilon_1}{\mathrm{d}\mu} \Big)
\end{equation}

\noindent where $R_\mathrm{Q} =h/e^2$ is the resistance quantum, $V = V_\mathrm{ac} + V_\mathrm{dc}$, and we have neglected contact resistance (see \cite{WeiSupp2018} for a derivation of Eq. 1, which takes contact resistance into account). We find that the conductance decreases at negative bias and increases at positive bias (Fig.\ 2D) -- indicating that $\varepsilon_1 > \varepsilon_2$ for both positive and negative bias. This implies that more magnons are absorbed at $\varepsilon_1$ than at $\varepsilon_2$. Because our contacts have all been fabricated identically, we conclude this is because $\varepsilon_1$ is closer to magnon generation than $\varepsilon_2$ (for both positive and negative bias, see Fig.\ 2, A and B). Using different sets of contacts and top gates (Fig.\ 2, E to H) we can change the relative distances of $\varepsilon_i$ to the locations of magnon generation. We confirm that for each configuration, the conductance values after $E_\mathrm{Z}$ correspond to a greater number of magnons absorbed at the site closer to magnon generation. 

This change to the conductance is not a consequence of QH breakdown. Conductance deviations after the Zeeman threshold that depend on the sign of $V_\mathrm{dc}$ are not explained by any current breakdown theories \cite{Nachtwei1999}. Additionally, we find that the threshold voltage bias does not depend on the lead configuration (Fig.\ 2), the size of the  $\nu = 1$ region (fig.\ S4), or the density of the $\nu = 1$ region (fig.\ S6) -- which is all inconsistent with trivial QH breakdown, but consistent with our magnon model. In total, we have measured this  $\nu = 1$ conductance deviation occurring at the Zeeman energy for eight devices of widely varying geometries (figs. S3, S4, and S11).  

	\begin{figure*}
	\includegraphics[width=0.9\textwidth]{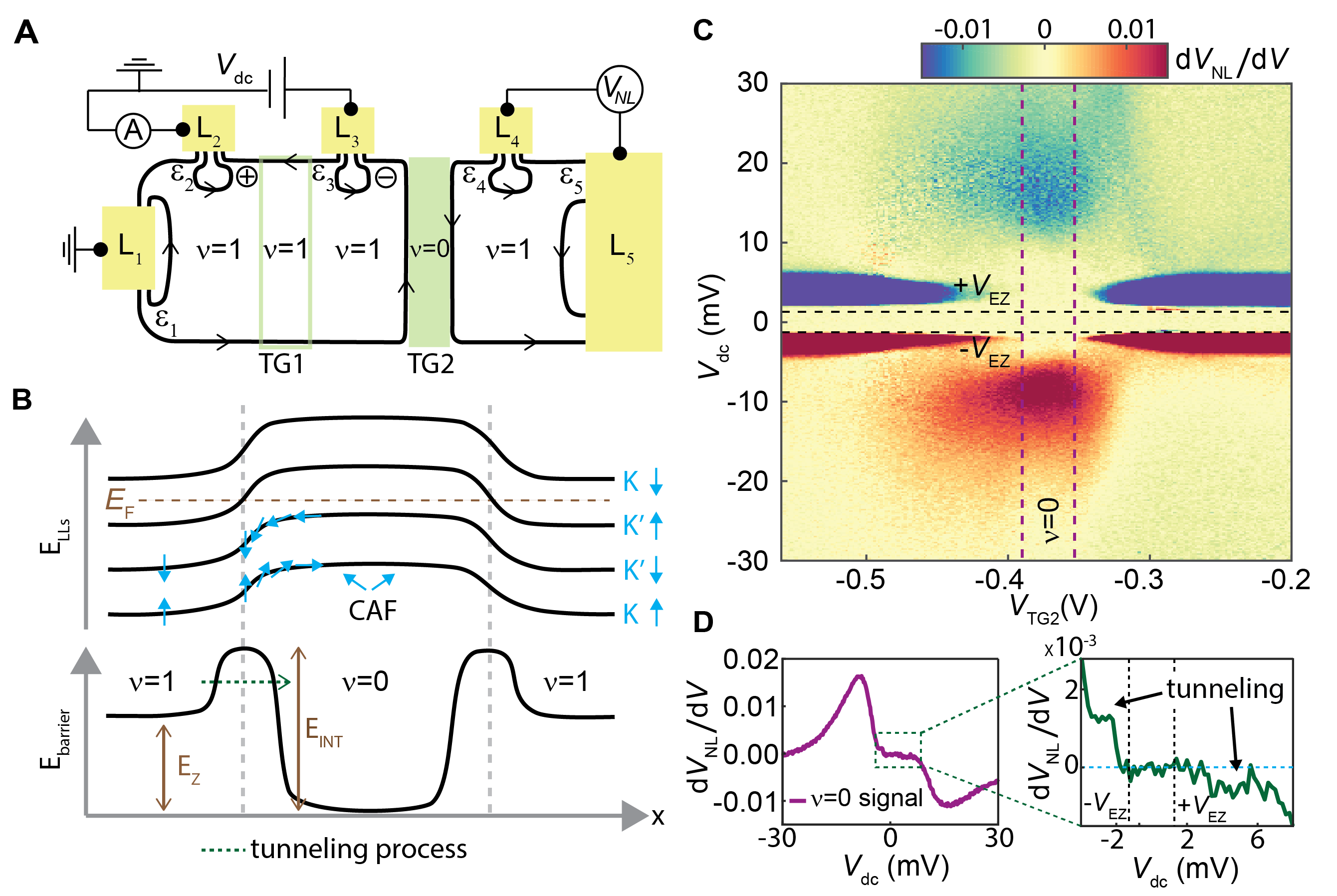}%
	\caption{\textbf{Non-local voltage signal due to magnon propagation through the $\nu = 0$ CAF.} \textbf{(A)} Schematic of the circuit used to measure $S_\mathrm{NL}$ in device 2 across a $\nu = 0$ region. $\nu_\mathrm{TG1} = 1$ for all measurements while $\nu_\mathrm{TG2}$ is swept from -1 to 1 ($\nu_{BG}$ = 1, $V_{BG}$ = 4 V).  \textbf{(B)} Top: Postulated spatial variation of the LLs and spin arrangement in a $\nu = 1/ \nu = 0 / \nu = 1$ geometry. Close to the interface between $\nu = 1$ and $\nu = 0$, spins in the two filled Landau levels prefer to be in an aligned antiferromagnetic (AF) arrangement. Deeper into the $\nu = 0$  region, the spins slowly rotate into the canted antiferromagnetic phase.  Because the minimum magnon energy in the aligned AF region is higher than $E_\mathrm{Z}$, it should present a barrier for incident magnons close to the energy threshold. Bottom: Energy barrier seen by the magnons as a function of position, where $E_\mathrm{INT}$ is the energy barrier of the interface. \textbf{(C)} When magnons are generated, we see another onset of $S_\mathrm{NL}$ at energies exceeding $\pm V_\mathrm{EZ}$ ($B$ = 8 T), indicating that higher energy magnons have overcome $E_\mathrm{INT}$ and have propagated through the $\nu_\mathrm{TG2}$ = 0 region. Purple dashed lines indicate a region where vertical line cuts were taken and averaged to obtain the line trace in (D). \textbf{(D)} A clear onset of $S_\mathrm{NL}$ is shown at biases exceeding $\pm V_\mathrm{EZ}$ when $\nu_\mathrm{TG2}$ = 0. It is not presently understood why the signal is asymmetric both in energy of onset and strength of signal. The zoomed-in region shows a clear increase in $S_\mathrm{NL}$ at -$V_\mathrm{EZ}$ and a signal consistent with a decrease, slightly offset from $+V_\mathrm{EZ}$, indicating that magnons can tunnel through the interface barrier at lower energies ($S_\mathrm{NL}$ is offset by 0.01 $\mu$V at $V_{dc}$ = 0 and is manually corrected for).}
	\label{fig:F2}
	\end{figure*} 

Thus far we have established that we are able to generate and absorb magnons at current carrying contacts. If these chargeless excitations propagate through the insulating bulk, we also expect to see signatures of magnon propagation and absorption via non-local voltage measurements (d$V_\mathrm{NL}$/d$V$—referred to as nonlocal signal $S_\mathrm{NL}$), away from the source-drain current. To measure $S_\mathrm{NL}$ we use L$_3$ and L$_2$ in device 2 as source-drain contacts, and use contacts L$_4$ and L$_5$ as voltage probes (Fig.\ 3A). These contacts are separated from the source-drain contacts by a top gate (TG2) which we tune between $\nu_\mathrm{TG2}$ = -2 and $\nu_\mathrm{TG2}$ = 2, where all other regions are tuned to $\nu=1$. The conductance between L$_3$ and L$_2$ drops at $V_\mathrm{EZ}$ in accordance with our model (Fig.\ 3B), whereas magnon generation is largely unaffected by TG2 (fig.\ S7, A). At $\nu_\mathrm{TG2}$ = 1 we measure a change in $S_\mathrm{NL}$ at $\pm V_\mathrm{EZ}$ due to the relative absorption at each magnon absorption site ($\varepsilon_i$).

The sign of S$_\mathrm{NL}$ indicates that there is more magnon absorption at sites closer to where magnon generation occurs. Through current conservation (\cite{WeiSupp2018}) we find that the measured differential voltage (unitless) is: 

\begin{equation} \label{eq2}
\frac{\mathrm{d} V_\mathrm{NL}}{\mathrm{d}V} = \Big( \frac{\mathrm{d} \varepsilon_4}{\mathrm{d} \mu} - \frac{\mathrm{d} \varepsilon_5}{\mathrm{d} \mu} \Big )
\end{equation}

The site labeled by $\varepsilon_4$ is closer to magnon generation than $\varepsilon_5$ for both negative and positive bias, so $|\mathrm{d}\varepsilon_4| > |\mathrm{d}\varepsilon_5|$. However, the differential change in voltage ($\mathrm{d}\varepsilon_i/\mathrm{d}\mu$) is negative for $V_\mathrm{dc} \geq V_\mathrm{EZ}$ and positive for $V_\mathrm{dc} \leq -V_\mathrm{EZ}$, corresponding to an overall negative value for $S_\mathrm{NL}$ at $V_\mathrm{dc} \geq V_\mathrm{EZ}$ and a positive value at $V_\mathrm{dc} \leq -V_\mathrm{EZ}$ (Fig. 3C). 

The device geometry used for our non-local measurements allows us to tune TG2 away from  $\nu_\mathrm{TG2} = 1 $, and thereby examine magnon transmission through different filling factors. We make two surprising observations. We observe that when $\nu_\mathrm{TG2} = -1 $ the signal $s_\mathrm{NL}$ is almost identical signal to when $\nu_\mathrm{TG2} = 1 $ (Fig.\ 3C and fig.\ S8). This signal arises in the absence of any charge leakage across the $\nu_\mathrm{TG2} = -1$ region (fig.\ S9), so that changes in $S_\mathrm{NL}$ can be attributed to magnon transport through the $\nu_\mathrm{TG2} = -1$ ferromagnet. This suggest that there is neither spin nor valley mismatch between the ferromagnetic states on either side of the boundary. We therefor propose an ordering of the LLs that does not require a spin or valley flip for magnons to travel across the interface between $\nu_\mathrm{BG} = 1$ and $\nu_\mathrm{TG2} = -1$ (Fig. 3D; see \cite{WeiSupp2018} for a theoretical discussion.)

In addition, we unexpectedly find that $S_\mathrm{NL}$ is suppressed at $\pm V_\mathrm{EZ}$ when $\nu_\mathrm{TG2} = 0$. For non-magnetic regions such as $\nu_\mathrm{TG2}=2$, it is expected that magnons will be blocked from passing through, as experimentally confirmed in Fig.\ 3C (the non-local signal occurring at the transition between $\nu = 1$ and $\nu =2$ is explained in Fig. S7E). However, $\nu = 0$ is purportedly a canted antiferromagnet which is theoretically capable of hosting even zero-energy magnons \cite{Takei2016}. It appears that the probability for an incident magnon to be transmitted across the junction between the $\nu = 0$ and $\nu = 1$ regions is very small for energies close to $E_\mathrm{Z}$. This may be caused by, in part, the mismatch in propagation velocities in the two phases, or a barrier due to the complex nature of the interface region. Close to the boundary with a $\nu = 1$ phase, the ground state of the $\nu = 0$ phase may not have canted spins but may instead be in an aligned antiferromagnet state, where spins are parallel to the magnetic field on one sublattice and antiparallel on the other. Eventually, far from the boundary, we may expect the local spin arrangement to rotate into the CAF orientation (Fig.\ 4B). In the transition region, the minimum magnon energy will be larger than $E_\mathrm{Z}$ due to effects of the valley-dependent interaction terms \cite{Kharitonov2012}, which were initially responsible for the antiferromagnet arrangement to be favored over the ferromagnetic arrangement. In order to cross from the $\nu = 1$ region to the CAF region, a magnon with energy close to $E_\mathrm{Z}$ would have to tunnel through the barrier region, and we would expect the transmission rate to be low. If the magnons have enough energy to overcome this barrier, they should be able to more easily enter the CAF region. Fig.\ 4C shows that we can experimentally exceed this barrier, where we see non-local signals at higher $|V_\mathrm{dc}|$ with signs in agreement with our magnon model. The onset of this magnon signal is unaffected by any charge transport across the $\nu_\mathrm{TG2} = 0$ region (fig.\ S10). Closely examining the signal at $\nu_\mathrm{TG2} = 0$, we see signals commencing at $\pm V_\mathrm{EZ}$ which we attribute to tunneling events across this $\nu_\mathrm{BG} = 1/\nu_\mathrm{TG2} = 0$ barrier (Fig.\ 4D).  

Note that all non-local signals (occurring at $\nu_\mathrm{TG2}$ =-1, 0, and 1) appear only in a finite band of $V_\mathrm{dc}$. This suppression of the differential voltage signal indicates that either magnon generation is suppressed, or alternatively, that the differently-spaced contacts begin to see identical amounts of magnon absorption once the system has reached a certain magnon density threshold. We further speculate that this cut-off could be related to the magnon bandwidth, but leave this to a future investigation. 

The experiments presented here introduce a method of using magnons to probe the SU(4) spin and valley anisotropies of graphene QH systems, whichc can be used to probe highly correlated states such as the fractional QH regime \cite{MacDonald1998}, or the quantum-spin Hall phase of monolayer graphene \cite{Young2014}. Owing to the theoretical prediction for spin superfluidity in the CAF state \cite{Takei2016}, this study paves the way for exploring and realizing dissipationless spin waves in a Bose-Einstein condensate (BEC) of magnons. Such condensates should result in a coherent precession of the spin in the QH magnet, which may be probed through emitted microwave radiation. Furthermore, coherent spin waves associated with a BEC may be able to propagate long distances with negligible dissipation, which could be tested by careful length dependence measurements.

	\section{Methods} 
	\subsection{Sample Fabrication} All devices consist of graphene encapsulated by two layers of hexagonal boron nitride (hBN) on doped Si chips with a 285 nm layer of SiO$_2$ that acts as a dielectric for the Si back gate. Graphene is mechanically exfoliated from bulk graphite obtained from NGS Naturgraphit GmbH using 1009R tape from Ultron Systems and subsequently encapsulated in hexagonal boron nitride (hBN) using a dry transfer process \cite{Wang2013}. Before the first metal deposition step, we annealed the devices in vacuum at $500^{\circ}$C to improve device quality. We then created top gates using electron-beam lithography and thermal evaporation of Cr/Au. We etched the devices into the desired geometry by reactive ion etching in $\mathrm{O}_2/\mathrm{CHF}_3$ using a PMMA/HSQ bilayer of resist (patterned by electron-beam lithography) as the etch mask. To fabricate edge-contacts to the graphene we etched through the entire hBN/graphene stack. We then created edge contacts by thermally evaporating Cr/Au while rotating the sample using a tilted rotation stage. 
	\subsection{Measurement} Our measurements were performed in a Leiden dry dilution refrigerator with a base temperature of 20 mK. Measurements of differential conductance were performed using a lock-in amplifier with an a.c. excitation voltage of 50 $\mu$V at 17.77 Hz. All measurements of differential conductance were corrected for contact/line resistances, which were independently determined by lining up the robust $\nu = 2$ QH conductance plateau with $2e^2/h$. 
	\bibliographystyle{naturemag}
	\bibliography{Bibliography}
	\section{Acknowledgements}
	\subsection{General} We thank A. H. Macdonald, J. D. Sanchez-Yamagishi,
	S. L. Tomarken, and S. P. Harvey for helpful discussions and
	feedback, X. Liu for fabrication help, and P. Kim for providing the
	transfer setup. \textbf{Funding}: Supported by the Gordon and Betty
	Moore Foundation’s EPiQS Initiative through grant GBMF4531;
	the U.S. Department of Energy, Basic Energy Sciences Office,
	Division of Materials Sciences and Engineering under award
	DE-SC0001819 (D.S.W. and T.v.d.S.); NSF Graduate Research
	Fellowship grant DGE1144152 (D.S.W.); the STC Center for
	Integrated Quantum Materials, NSF grant DMR-1231319 (B.I.H.);
	and the Elemental Strategy Initiative conducted by MEXT,
	Japan, and JSPS KAKENHI grant JP15K21722 (K.W. and T.T.).
	Nanofabrication was performed at the Center for Nanoscale
	Systems at Harvard, supported in part by NSF NNIN award
	ECS-00335765. \textbf{Author contributions}: D.S.W., T.v.d.S., B.I.H.,
	and A.Y. conceived and designed the experiments; D.S.W.
	fabricated the devices; D.S.W. and A.Y. performed the
	experiments; D.S.W., T.v.d.S., S.H.L., B.I.H., and A.Y. analyzed
	the data and wrote the paper; and K.W. and T.T. synthesized the
	hexagonal boron nitride crystals. \textbf{Competing interests}: The
	authors declare no competing financial interests. \textbf{Data and
	materials availability}: All measured data are available in the
	supplementary materials.

	\section{Correspondence} Correspondence and requests for materials should be addressed to A.Y. (email: yacoby@physics.harvard.edu).	
	\newcommand{\beginsupplement}{%
		\setcounter{table}{0}
		\renewcommand{\thetable}{S\arabic{table}}%
		\setcounter{figure}{0}
		\renewcommand{\thefigure}{S\arabic{figure}}%
		\setcounter{equation}{0}
		\renewcommand{\theequation}{S\arabic{equation}}
	}
		\beginsupplement
		
		\section{Supplementary Information}
		
		
		\noindent \textbf{Supplementary Note 1. Equilibration of edge states at ‘hot spots’}\\
		
		One question that arises from this study is why, after decades of experimental investigation into QH ferromagnets, has this phenomenon not been observed in GaAs quantum wells? We posit that this is due to the readiness of edge states in graphene to equilibrate over small length scales due to the sharp confining potentials. Because there is a limited spatial range over which the `hot spot' magnon generation can occur adjacent to the $\nu = 1$ ferromagnetic bulk, spin-flip induced edge equilibration must occur over short lengths. Past studies have shown that in graphene edges of the same spin are able to fully equilibrate over length scales $< 1$ $\mu$m \cite{Amet2014, Williams2007,Ozyilmaz2007}, while similar studies done in GaAs found typical lengths of around tens of microns, and sometimes up to 200 $\mu$m \cite{VanWees1991, Alphenaar1990, Komiyama1989}. In order to experimentally verify that magnons are generated at these corner `hot spots', we have fabricated a device with gated corners showing conductance changes at $E_\mathrm{Z}$ in accordance with our magnon model (Fig.\ S11). 
		
		The difference in experimentally-determined equilibration lengths between graphene and GaAs is likely due to the sharper confining potentials in graphene, which allow for small spatial edge channel separation—increasing the likelihood of inter-channel scattering. Additionally, a smooth potential may allow for edge reconstruction \cite{Chklovskii1992}, which if present, could also affect the inter-channel scattering rate and limit magnon generation. Experiments in GaAs based systems have shown that edge recondsturction plays an important role due to the smooth confining potential \cite{Ahlswede2002}. The graphene devices investigated in this work have gate electrodes located at just a few tens of nanometers distance away, likely limiting the amount of edge reconstruction. Furthermore, the close proximity of the metal gates to the graphene may screen the electric fields that cause edge reconstruction, which is another potential difference with GaAs-based systems \cite{Li2013}.
		
		However, interestingly, we note that although an electrostatically-defined confinement potential is able to suppress the magnon signal, it does not eliminate it completely (Fig. S11). This suggests that strong edge disorder is not required for the $\nu$ = 2 and $\nu$ = 1 edge channels to equilibrate, and that it is possible that magnons could be generated in GaAs devices with a sharp electrostatic confinement potential. We note, however, that while magnon generation may be possible, magnon propagation may not be as efficient in GaAs systems due to large spin-orbit coupling \cite{Muller1992} and more nuclear spins \cite{Burkard1999} relative to graphene, which could facilitate magnon dissipation. Such dissipative processes would be important because, as we describe in the main text, magnon generation itself does not affect the sample conductance -- only when magnons are able to propagate and are absorbed in by electrons in other edge channels do we detect a change in sample conductance. 
		
		Additionally, we note that even when we do not explicitly add an extra edge state near the contacts (by gating the side regions to $\nu$  = 2), contact doping of the graphene by the Cr/Au leads \cite{Giovannetti2008} introduces additional spin-down edge states—which also leads to magnon generation at EZ.  \\
		
		
		\noindent \textbf{Supplementary Note 2. Calculating $V_\mathrm{dc}$ necessary to exceed $V_\mathrm{EZ}$ given a finite contact resistance}\\
	
		 In a two-terminal measurement, the applied bias voltage ($V_\mathrm{dc}$) drops over both the contact resistances at both the source and the drain. The d.c. current is therefore
		
		\begin{equation}\label{eqS1.1}
		I_\mathrm{dc} = \frac{V_\mathrm{dc}}{2R_\mathrm{C}+R_\mathrm{Q}} 
		\end{equation}	
		
		\noindent where $R_\mathrm{Q}$ is the quantum resistance of an edge channel, and where $R_\mathrm{C}$ includs both the contact resistance at each lead as well as the filtering on the lines. The filtering on each line is 4.5k$\Omega$, and the contact resistance at each lead of a typical device is about 500$\Omega$.  
		 
		The actual d.c. voltage that drops over the edge channel ($V_\mathrm{dc}'$) is therefore given by
		
		\begin{equation}\label{eqS1.2}
		V_\mathrm{dc}' = I_\mathrm{dc} R_\mathrm{Q}
		\end{equation}	
		
		\noindent Therefore:
		
		\begin{equation}\label{eqS1.3}
		V_\mathrm{dc} = \frac{V_\mathrm{dc}' \Big( 2R_\mathrm{C}+R_\mathrm{Q} \Big)}{R_\mathrm{Q}}
		\end{equation}	
		
	 In our figures, we use `$V_\mathrm{EZ}$' ($V_\mathrm{EZ}$ = -$E_\mathrm{Z}/e$) to denote the bias at which -$eV_\mathrm{dc}'$ reaches the Zeeman energy ($E_\mathrm{Z} = g \mu_B B_T$). \\
		
		
		\noindent \textbf{Supplementary Note 3. Circuit analysis for two-terminal conductance measurement.} \\
		
		
		\twocolumngrid
		\begin{figure}[h]
			\includegraphics[width=0.5\textwidth]{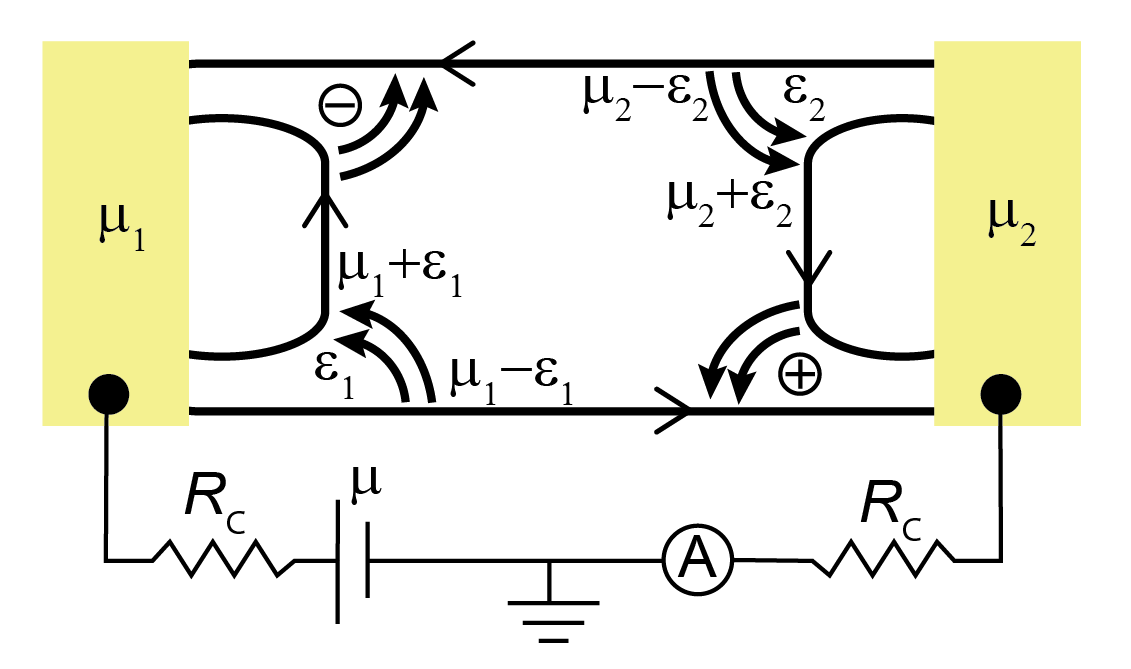}
			\caption{\textbf{Schematic of a two-terminal device where a voltage is sourced at the left contact and drained at the right.} The negative and positive signs denote magnon generation for negative and positive $V_\mathrm{dc}$ respectively. $\varepsilon_1$ and $\varepsilon_2$ denote locations where magnon absorption occurs. Arrows indicate how the chemical potential redistributes after magnon generation, and the chemical potential of the edge states after magnons are absorbed are labeled. the electrochemical potential applied by the voltage source is defined as $\mu$. $\mu_1$ and $\mu_2$ are the chemical potential reservoirs connected to the source and drain via a contact resistance $R_\mathrm{C}$ (assumed to be identical for both contacts).}
			\label{fig:Supp1}
		\end{figure}
		\twocolumngrid

		In Fig. S1 The (particle) current conservation equation at the source reservoir (labeled $\mu_1$) is:
		
		\begin{equation}\label{eqS2.1}
		\frac{2 \mu_1}{R_\mathrm{Q}} = \frac{\mu_2 -\varepsilon_2}{R_\mathrm{Q}} + \frac{\mu_1 +\varepsilon_1}{R_\mathrm{Q}} + \frac{\mu -\mu_1}{R_\mathrm{C}}
		\end{equation}
		
		\noindent where $R_\mathrm{Q}$ is the resistance quantum. As described in the main text, $\varepsilon_i$  denotes the chemical potential redistributed between edge states at the $i$th contact. Additionally, although there are indeed also spin-flips occurring at the negative-bias magnon generation location when positive bias magnons are being generated (and for the reverse case), we ignore these in our analysis because they do not contribute to changes in the conductance.
		
		The equation at the drain contact is:
			
		\begin{equation}\label{eqS2.2}
		\frac{\mu_2 +\varepsilon_2}{R_\mathrm{Q}} + \frac{\mu_1 -\varepsilon_1}{R_\mathrm{Q}} = \frac{2\mu_2}{R_\mathrm{Q}} + \frac{\mu_2}{R_\mathrm{C}}
		\end{equation}
		
		\noindent Solving for $\mu_2$ we find
		
		\begin{equation}\label{eqS2.3}
		\mu_2 = \frac{\mu +\varepsilon_2 - \varepsilon_1}{2+\frac{R_\mathrm{Q}}{R_\mathrm{C}}}
		\end{equation}

		Using the chemical potential of the voltage source as $\mu = -eV_\mathrm{dc}$, and the charge current $I = -eI_\mathrm{P}$ (where $I_\mathrm{P}$ is defined as $\frac{1}{e^2}$ $\frac{\mu_2}{R_\mathrm{C}}$  to normalize the units) the differential conductance measured is
		
		\begin{equation} \label{eqS2.4}
		\begin{split}
		\frac{\mathrm{d}I}{\mathrm{d}V_\mathrm{dc}} & = \frac{\mathrm{d}I_\mathrm{P}}{\mathrm{d}\mu}  = \frac{\big(\mathrm{d}\mu_2/R_\mathrm{C} \big)}{\mathrm{d}\mu}   
		\end{split}
		\end{equation}

		\begin{equation} \label{eqS2.4}
		\begin{split}
		\frac{\mathrm{d}I}{\mathrm{d}V_\mathrm{dc}} = \frac{1}{2 R_\mathrm{C}+R_\mathrm{Q}} \Big( 1+\frac{\mathrm{d}\varepsilon_2}{\mathrm{d}\mu}-\frac{\mathrm{d}\varepsilon_1}{\mathrm{d}\mu} \Big)
		\end{split}
		\end{equation}
		
		\noindent This becomes Equation 1 (main text) in the absence of contact resistance ($R_\mathrm{C}$ =0).\\
		

		\noindent\textbf{Supplementary Note 4. Circuit analysis for non-local voltage measurements.}\\
		

  	\twocolumngrid
			\begin{figure}[h]
			\includegraphics[width=0.5\textwidth]{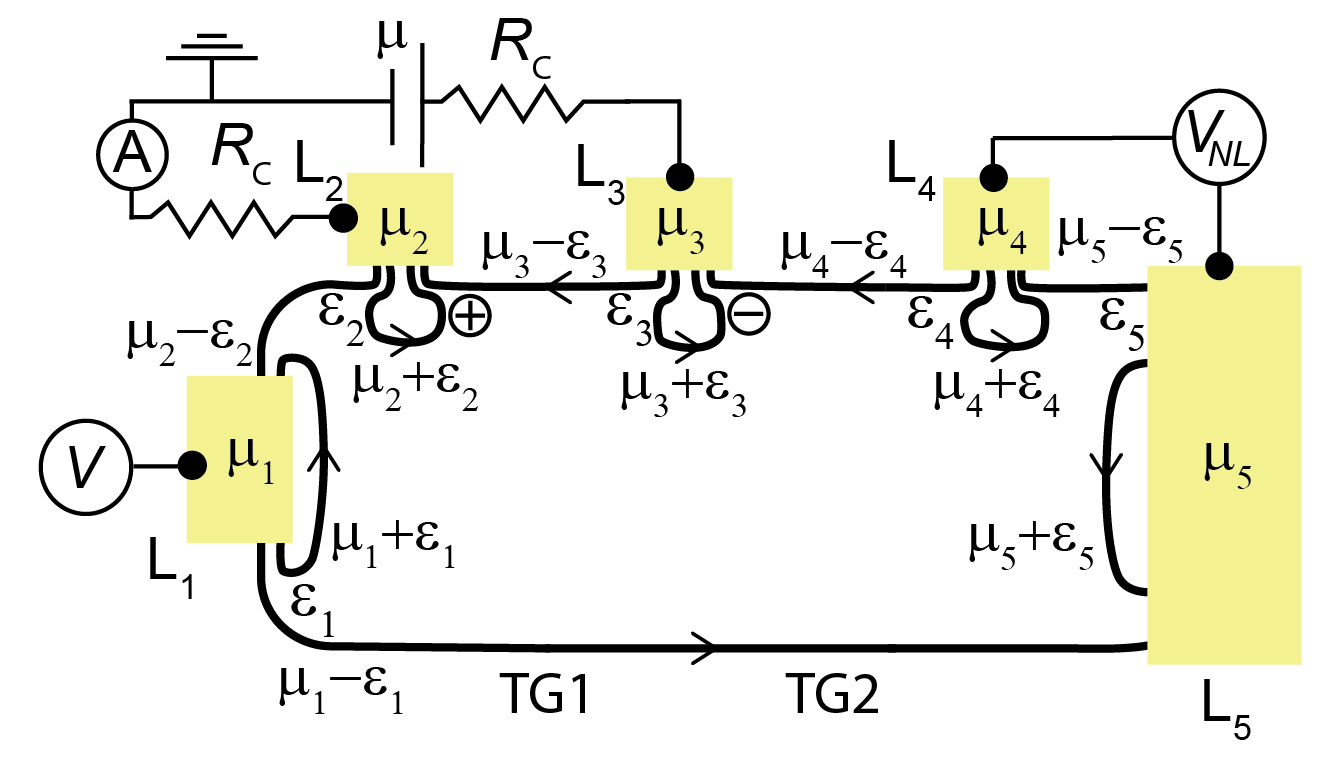}
			\caption{\textbf{Schematic circuit diagram of the multi-terminal device} Device 2 - optical micrograph shown in Fig.\ 2A) that is used to measure $S_\mathrm{NL}$. $\mu_1$, $\mu_2$, $\mu_3$, $\mu_4$ and $\mu_5$ are the chemical potential reservoirs connected to each contact (L$_1$-L$_5$) by a contact resistance $R_\mathrm{C}$. The electrochemical potential applied by the voltage source is defined as $\mu$. Voltage is sourced at L$_3$ and drained from L$_2$. L$_1$ is floating. $S_\mathrm{NL}$ is measured between L$_4$ and L$_5$. The negative and positive signs denote magnon generation for negative and positive $V_\mathrm{dc}$ respectively. $\varepsilon_1$, $\varepsilon_2$, $\varepsilon_3$, $\varepsilon_4$, and $\varepsilon_5$ label locations of magnon absorption.  }
			\label{fig:Supp1}
		\end{figure}
	\twocolumngrid

		 In Fig. S2 the chemical potential of the edge states after magnons are absorbed are labeled as $\mu_{i} - \varepsilon_{i}$ and $\mu_i +\varepsilon_{i}$ where `$i$' denotes the contact where the chemical potential originates.  We calculate the conductance expected after the Zeeman energy has been reached. This device has 5 contacts in total. We write a current conservation equation at each contact:
		
		\begin{equation}\label{eqS3.1}
		\mathrm{L}_1: \mu_2 - \varepsilon_2+\mu_1 + \varepsilon_1 = 2 \mu_1
		\end{equation}
		
	    \begin{equation}\label{eqS3.2}
		\mathrm{L}_2: \mu_3 - \varepsilon_3+\mu_2 + \varepsilon_2 = 2 \mu_2 + \mu_2 \frac{R_\mathrm{Q}}{R_\mathrm{C}}
		\end{equation}
		
		\begin{equation}\label{eqS3.3}
		\mathrm{L}_3: (\mu - \mu_3) \frac{R_\mathrm{Q}}{R_\mathrm{C}} + \mu_3 + \varepsilon_3 + \mu_4 - \varepsilon_4 = 2 \mu_3
		\end{equation}
		
		\begin{equation}\label{eqS3.4}
		\mathrm{L}_4: \mu_5 - \varepsilon_5+\mu_4 + \varepsilon_4 = 2 \mu_4
		\end{equation}
		
		\begin{equation}\label{eqS3.5}
		\mathrm{L}_5: \mu_1 - \varepsilon_1+\mu_5 + \varepsilon_5 = 2 \mu_5
		\end{equation}

		\noindent Solving for $\mu_2$ we find
	
		\begin{equation}\label{eqS3.6}
		\mu_2 = \frac{\mu +\varepsilon_2 - \varepsilon_3}{2+\frac{R_\mathrm{Q}}{R_\mathrm{C}}}
		\end{equation}

		\noindent Therefore,
	
		\begin{equation} \label{eqS3.7}
		\begin{split}
		\frac{\mathrm{d}I}{\mathrm{d}V_\mathrm{dc}}  = \frac{\mathrm{d}I_\mathrm{P}}{\mathrm{d}\mu} = \frac{\mathrm{d}\big(\mu_2/R_\mathrm{C} \big)}{\mathrm{d}\mu} = \frac{1}{2 R_\mathrm{C}+R_\mathrm{Q}} \Big( 1+\frac{\mathrm{d}\varepsilon_2}{\mathrm{d}\mu}-\frac{\mathrm{d}\varepsilon_3}{\mathrm{d}\mu} \Big)
		\end{split}
		\end{equation}

		\noindent The non-local voltage measured is:
		\begin{equation}\label{eqS3.8}
		S_\mathrm{NL} = \frac{\mathrm{d}V_\mathrm{NL}}{\mathrm{d}V} = \Big( \frac{\mathrm{d}\varepsilon_4}{\mathrm{d}\mu}-\frac{\mathrm{d}\varepsilon_5}{\mathrm{d}\mu} \Big)
		\end{equation}

		\noindent By defining $S_\mathrm{NL}$ as the difference between two voltage probes, any edge current which reaches the two voltage probes should not affect the measurement --- although we do see some small background voltage which is explained in Fig.\ S7, B.
		
		A similar circuit analysis can be done for any of the configurations found in the main text or supplementary materials. \\ \\


	\noindent\textbf{Supplementary Note 5. Theoretical Notes.}\\
	
	We first note that the energy levels shown in the Figure 3D are only schematic. The actual
	Landau levels will be broadened due to electron-electron interactions and, perhaps, disorder. The curves represent more accurately the energy in the middle of the Landau level, and the ordering of the levels is more meaningful than the actual energies.
	
	Our ordering of levels was guided by the following observations. For a uniform graphene
	system at $\nu$ = 0, it is believed that the valley anisotropy energy is large compared to the Zeeman energy, and that the ground state is a canted antiferromagnet state \cite{Young2014, Kharitonov2012}. In this half-filled N=0 Landau level, there is one electron per flux quantum on each sublattice, with spins oriented predominantly in opposite directions. In the absence of Zeeman coupling the antiferromagnetic axis could point equally well in any direction, with no difference in the energy \cite{Jung2009}. In the presence of the Zeeman field, there is a small energy gain for the antiferromagnetic axis to line up in the x-y plane, perpendicular to the Zeeman field, allowing the spins on both sublattices to cant slightly in the direction of the Zeeman field. The energy gain for this is of order $E_Z^2$/$E_A$, where $E_A$ is the valley anisotropy energy.
	
	For a general filling fraction in the range $-1 < \nu < 0$, if one calculates the ground state energy in a restricted Hartree-Fock approximation, which assumes that only two of the possible spin-valley states are occupied by electrons, one generally finds that one valley, say the K valley, has one electron per flux quantum, while the other valley has occupancy $1 + \nu < 1$. For fillings very close to $\nu$ = 0, the system may remain in a canted configuration, but for $|\nu|$ exceeding a critical value, of order $E_Z/E_A$, it will be more favorable for the antiferromagnetic axis to align in the z-direction, so that the majority spin is fully aligned with the Zeeman field. (See, e.g. the discussion in \cite{Abanin2013}) Similarly, for $0 < \nu < 1$, we would find the antiferromagnetic spin axis to be aligned with the magnetic field, except for a small region close to $\nu = 0.$
	
	In a situation where the electron density varies rapidly in space, the spin and valley orientationsshould be determined by the dominant exchange energy, arising from the long-range part of theCoulomb interaction, which is indifferent to the specific orientation of the occupied levels in spin-valley space, but disfavors any rapid changes or discontinuities in the occupations. In a	boundary between $\nu =-1$ and $\nu = 1$, we are forced to have two discontinuities in occupancy, but we can avoid any other discontinuities, if we choose to fill the levels in the order suggested in	Fig. 3D. Moreover, it is likely that in a relatively steep boundary, the canted orientation will be
	completely suppressed, and that spins will remain quantized along the z-axis. We have seen in a previous study an absence of mixing between spin states at a $\nu = 1/ \nu = -1$ interface, supporting our assumption of spin alignment in the present case \cite{Wei2017}. 
	
	By contrast, when the filling fraction is $\nu = 0$ under the center of our gate, it is likely that the system will assume the canted orientation near the center of the gate. At the same time, there should be a strip on either side of the gate, where the filling fraction is intermediate between $\nu = 1$ and  $\nu = 0$, where the antiferromagnetic axis is in the z-direction. An interval where the filling fraction is between  $\nu = 1$  and  $\nu = -1$ , with spin axes parallel to z, will act as a barrier, to a spin wave incident from a region where  $\nu = 1$, as the energy at the bottom of the spin wave band will be raised by an amount of order the valley anisotropy energy (This should be small compared to the Coulomb exchange energy, but larger than the Zeeman energy) \cite{Pientka2017}. In the case where the filling under the gate is $\nu = -1$, we would expect the barrier regions at the two sides to be relatively thin, and it is plausible that the spin waves can tunnel rather easily through the barrier region. When the filling at the gate center is ν = 0, we would expect the barriers to be much thicker, and tunneling through the barriers should be reduced accordingly.
	
	In a bulk region where the filling is very close to ν = 1, we expect that the unoccupied spin state will have its spin opposite to the magnetic field, but it will have no particular preference for either the K or K’ valley or an arbitrary linear combination of them. Different valley polarizations may be selected near the physical boundaries of the sample, but we expect that the valley orientation in the vicinity of a gate where the charge density varies rapidly should be determined by energy considerations under the gate. It should cost relatively little energy for the valley orientation to vary smoothly between the sample edges and the gate, and we would not expect spin-wave propagation to be affected by such variations.
	
	Our analysis, based on a Hartree-Fock approximation, ignores correlation effects, which can lead to fractional quantized Hall states, varying spin polarization, and transitions between different spin states in uniform graphene sample \cite{Abanin2013, Feldman2013}. However, we would not expect such correlation effects to be important in the present case, where the charge density varies considerably on a sub-micron scale. 
	
		\begin{figure*}[t]
			\includegraphics[width=0.7\textwidth]{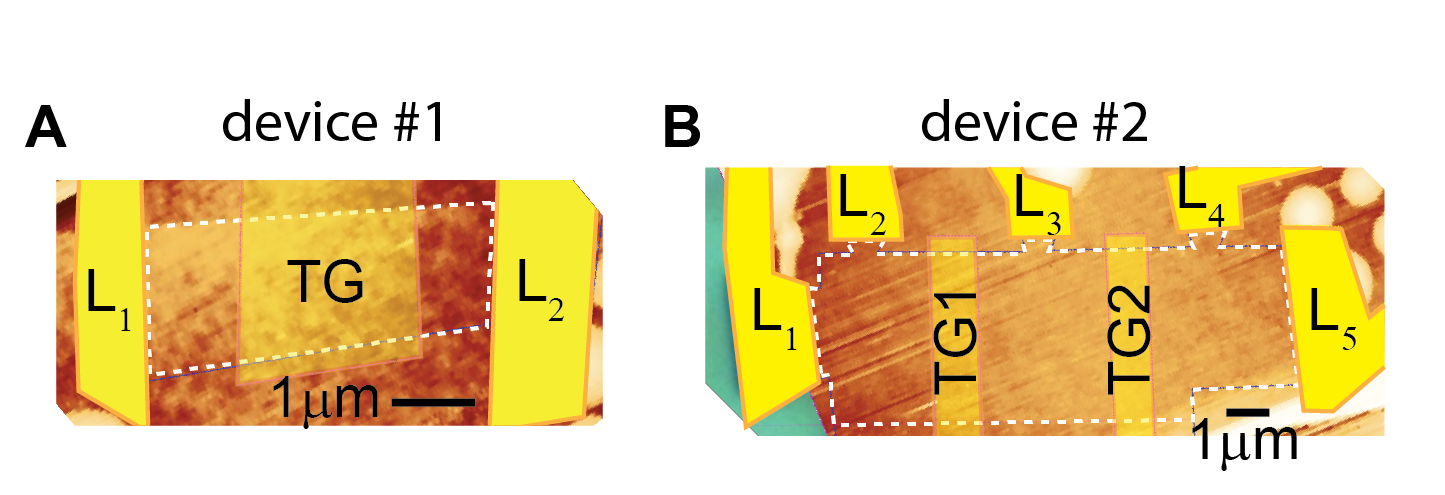}
			\caption{\textbf{Atomic force microscope (AFM) images of device 1 and device 2} \textbf{(A-B)} AFM images of the hBN/graphene/hBN heterostructures used for device 1 and device 2. Leads are illustrated in solid yellow and top gates are in transparent yellow. Dashed white lines outline the graphene flake. Scale bar: 1 $\mu$m. }
			\label{fig:Supp1}
		\end{figure*}

		\begin{figure*}[t]
			\includegraphics[width=0.8\textwidth]{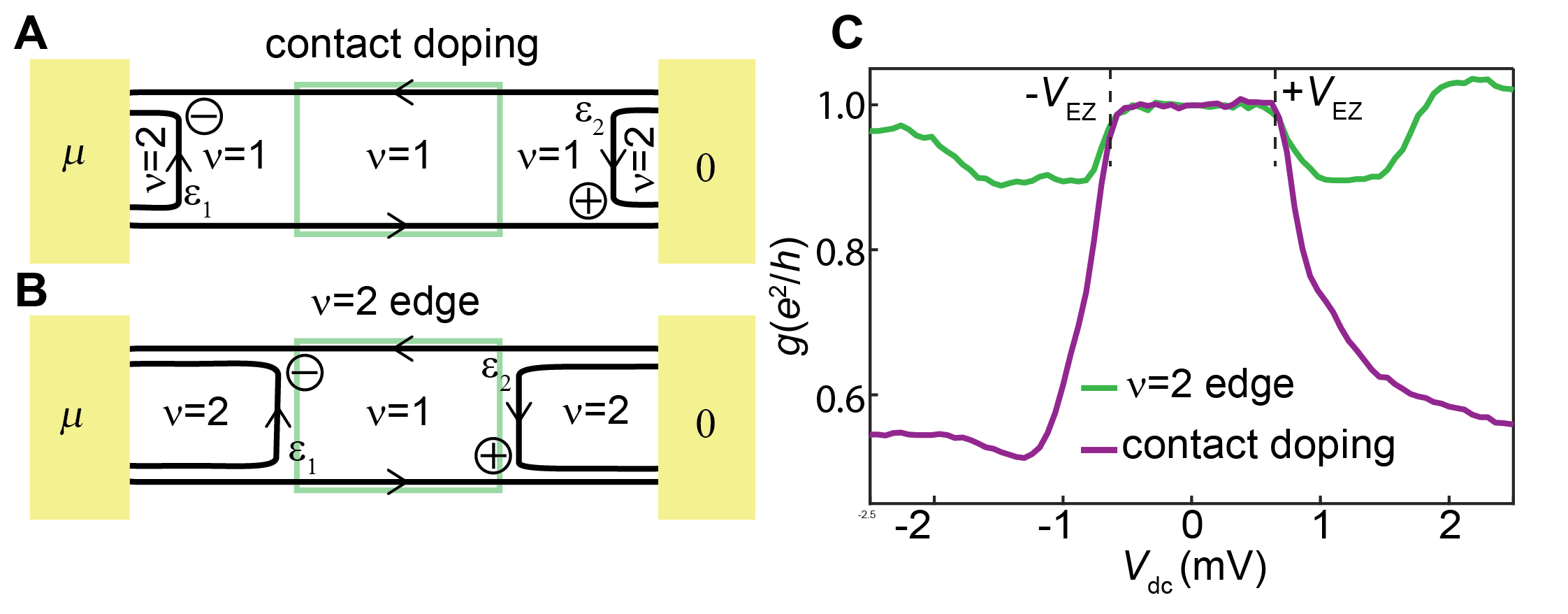}
			\caption{\textbf{Comparison of spin-reservoirs from contact doping and spin reservoirs from the $\nu=2$ edge (device 1).} \textbf{(A)} Schematic of device 1 where both the top and back-gated regions are set to $\nu=1$ and the magnon generation and absorption occurs at the contacts (See Fig.\ 1 for optical micrograph, and Fig.\ S3 for AFM image). The chemical potential redistribution at each magnon absorption site `$i$' is labeled by $\varepsilon_i$ (see discussion of $\varepsilon_i$ in the main text). \textbf{(B)} Schematic where the top-gated region is set to $\nu=1$ and the back-gated regions are set to $\nu=2$. Magnon generation and absorption occurs at the interface between $\nu=1$ and $\nu=2$. \textbf{(C)} Two-terminal conductance measurement at $B$ = 4 T where a constant d.c. voltage ($V_\mathrm{dc}$) and a 50 $\mu$V a.c. excitation voltage ($V_\mathrm{ac}$) are applied to the left contact and the differential conductance (d$I$/d$V$, where V = $V_\mathrm{dc}$+$V_\mathrm{ac}$) is measured through the right contact. The two cases are compared, one in which contact doping provides an opposite-spin reservoir as shown in (A) ($V_{BG}$ =1.24 V and $V_{TG}$= 0.12 V) – the other where the $\nu$=2 provides the opposite-spin reservoir as shown in (B) ($V_{BG}$ =3 V and $V_{TG}$ = -0.18V. The breakdown of the quantized $\nu = 1$ plateau occurs at identical values of $\pm V_\mathrm{dc}$, although the value of the conductance decrease changes dramatically. This is likely due to the changes to the relative magnitudes of magnon absorption at the different corners. The magnitudes of $\varepsilon_1$ and $\varepsilon_2$ may change as the $\nu = 1$ area changes. There may also be some effects on the magnitudes of $\varepsilon_1$ and $\varepsilon_2$ that arise from changing the nature of the spin reservoirs.}
			\label{fig:Supp2}
		\end{figure*}

		\begin{figure*}[t]
		\includegraphics[width=0.7\textwidth]{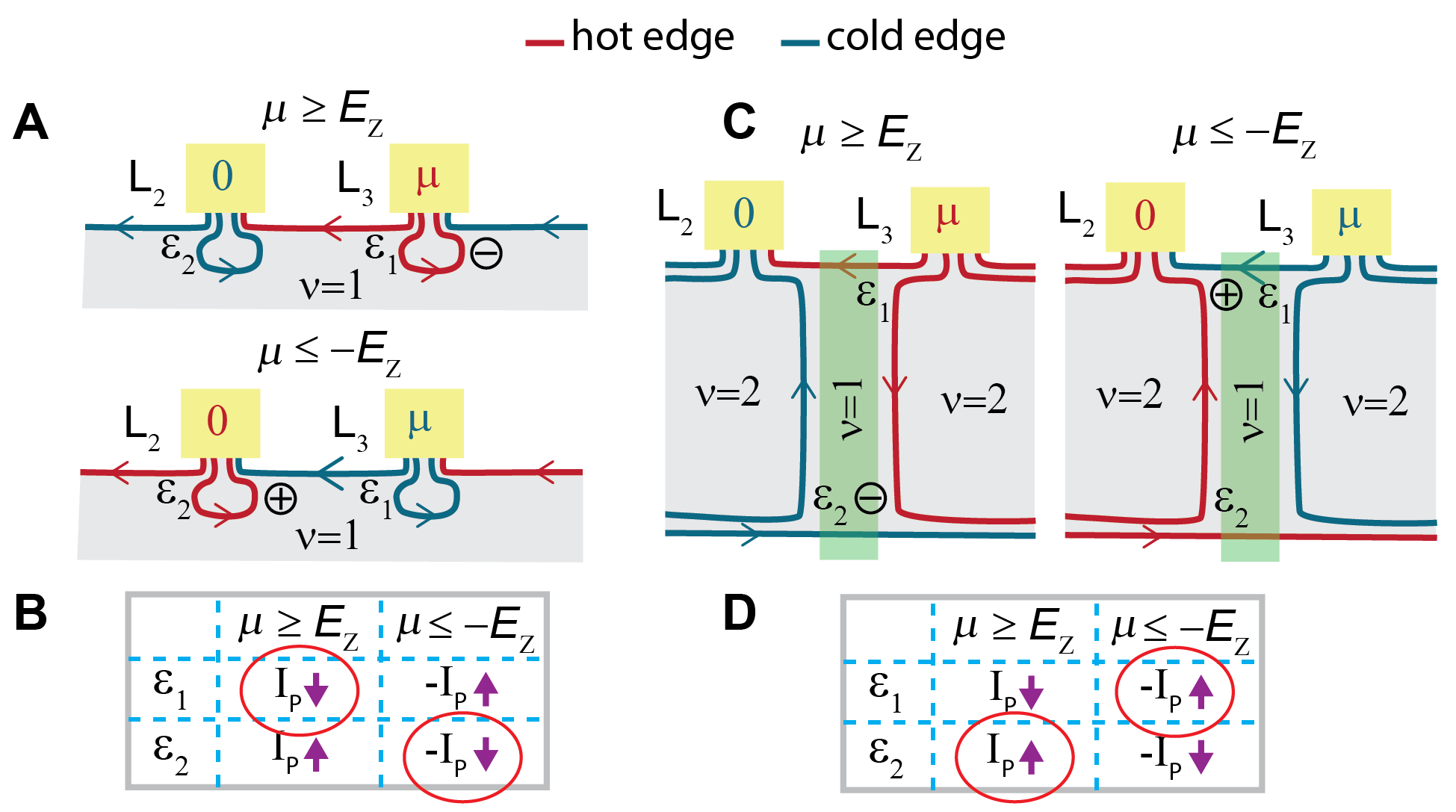}
		\caption{\textbf{Effect of relative magnon absorption on conductance using different lead configurations.} Schematic illustrations of different two-terminal conductance measurement using leads L$_3$ and L$_2$ where hot (cold) edges are colored red (blue), for both $\mu \geq E_\mathrm{Z}$ and $\mu \leq -E_\mathrm{Z}$ . The magnon generation site is labeled by the plus (minus) sign for positive (negative) bias (See Fig.\ 2A for optical micrograph, and Fig. S3 for AFM image). \textbf{(A)} Measurement where the entire device is tuned to $\nu =1$, so TG1 is not shown. (Upper panel) $\mu \geq E_\mathrm{Z}$ $(-eV_\mathrm{dc})$: magnon absorption at  $\varepsilon_1$ transfers chemical potential from a forward moving edge to a backwards moving edge --- causing the particle current ($I_\mathrm{P}$ where $I_\mathrm{P}$  = -$I/e$) to decrease. Conversely, magnon absorption at $\varepsilon_2$ transfers chemical potential from a backward moving edge to a forward moving edge, increasing $I_\mathrm{P}$. (Lower panel) $\mu \leq -E_\mathrm{Z}$ : magnon absorption at  $\varepsilon_1$ ($\varepsilon_2$) causes an increase (decrease) in $|-I_\mathrm{P}|$. \textbf{(B)} A summary of the effects of  $\varepsilon_1$ and  $\varepsilon_2$ at $\mu \geq E_\mathrm{Z}$ and $\mu \leq -E_\mathrm{Z}$. For $\mu \geq E_\mathrm{Z}$,  $\varepsilon_1$ is closer to the magnon generation site, so the current change caused by  $\varepsilon_1$ is predicted to be dominant and is circled in red. For $\mu \leq -E_\mathrm{Z}$, $\varepsilon_2$  is closer to the magnon generation site, so the current change caused by $\varepsilon_2$ is predicted to be dominant and is circled in red. \textbf{(C)} Measurement where the region under TG1 ($\nu_\mathrm{TG1}$) is tuned to $\nu_\mathrm{TG1}$ = 1 while the regions outside, tuned by the back gate ($\nu_\mathrm{BG}$), are set to $\nu_\mathrm{BG}$ = 2, providing a spin-down reservoir in the inner edge channel. (Left panel) $\mu \geq E_\mathrm{Z}$ ($-eV_\mathrm{dc}$): magnon absorption at $\varepsilon_1$ transfers chemical potential from a forward moving edge to a backwards moving edge --- causing the particle current ($I_\mathrm{P}$ where $I_\mathrm{P}$ = $-I/e$) to decrease. Conversely, magnon absorption at $\varepsilon_2$ transfers chemical potential from a backward moving edge to a forward moving edge, increasing $I_\mathrm{P}$. (Right panel) $\mu \leq -E_\mathrm{Z}$ : magnon absorption at  $\varepsilon_1$  ($\varepsilon_2$) causes an increase (decrease) in $|-I_\mathrm{P}|$. \textbf{(D)} A summary of the effects of  $\varepsilon_1$ and $\varepsilon_2$ at $\mu \geq E_\mathrm{Z}$ and $\mu \leq -E_\mathrm{Z}$. For $\mu \geq E_\mathrm{Z}$, $\varepsilon_2$ is closer to the magnon generation site, so the current change caused by $\varepsilon_2$ is predicted to be dominant and is circled in red. For $\mu \leq -E_\mathrm{Z}$,  $\varepsilon_1$  is closer to the magnon generation site, so the current change caused by $\varepsilon_1$ is predicted to be dominant and is circled in red.}
		\label{fig:Supp2}
	\end{figure*}

	\begin{figure*}[t]
		\includegraphics[width=0.5\textwidth]{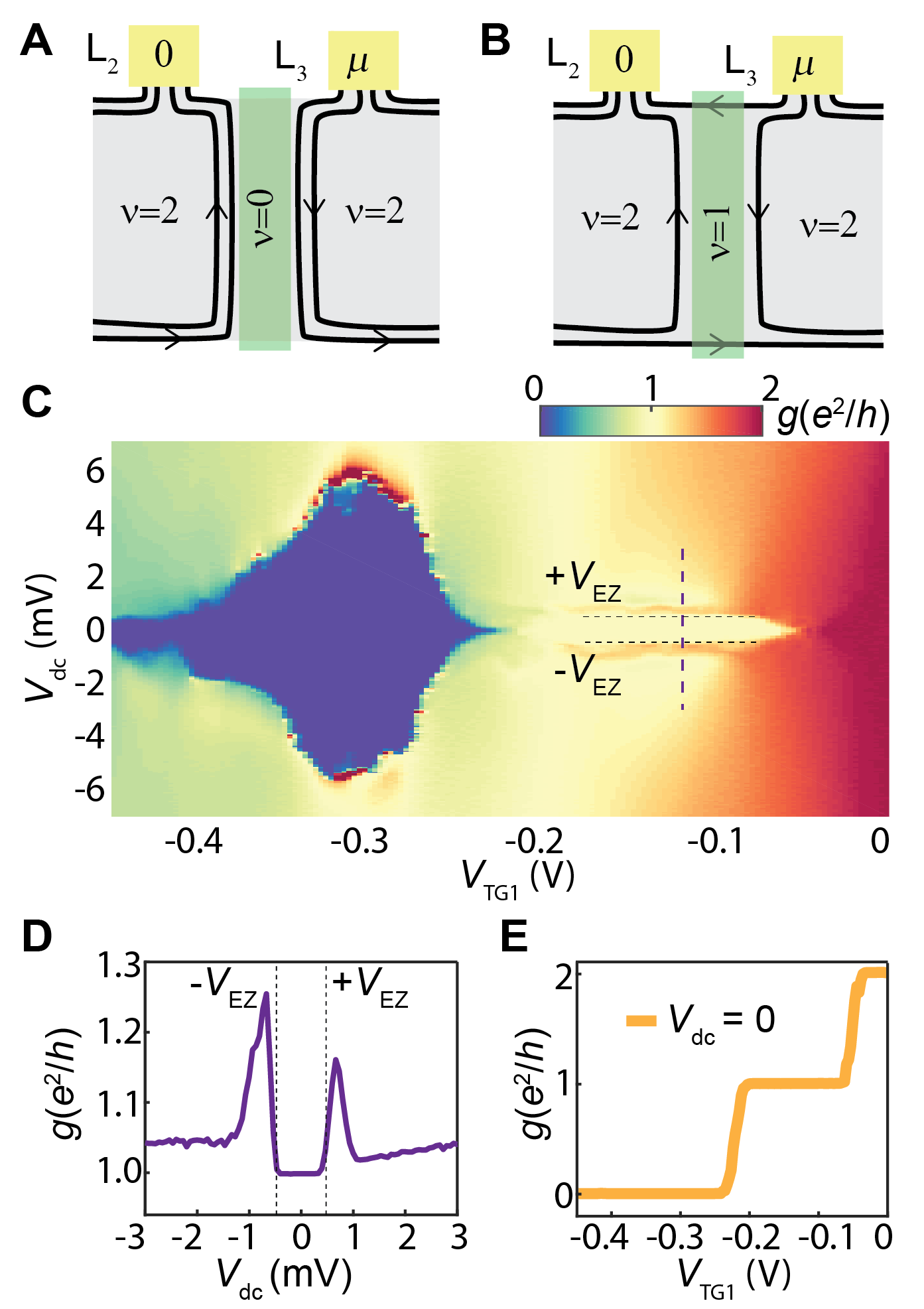}
		\caption{\textbf{Comparison of the breakdown of $\nu=0$ and $\nu=1$ LLs as a function of density (device 2).} \textbf{(A)} The region under top gate 1 (TG1) is at $\nu_\mathrm{TG1}=0$ while outside regions, gated only by the back gate, are at $\nu_\mathrm{BG}=2$. \textbf{(B)} $\nu_\mathrm{TG1}=1$ and $\nu_\mathrm{BG}=2$. \textbf{(C)} Two-terminal conductance measurement at $B$ = 3 T where a constant d.c. voltage ($V_\mathrm{dc}$) and a 50 $\mu V$ a.c. excitation voltage are applied to L$_3$ and the differential conductance (d$I$/d$V$) is measured through L$_2$. TG1 is swept from $\nu_\mathrm{TG1}=0$ to $\nu_\mathrm{TG1}=2$, and $\nu_BG = 2 $($V_BG$ = 1.8V). The horizontal black dashed lines denote $\pm V_\mathrm{EZ}$ and the location of the line cut taken in (D) is shown by the vertical purple dashed line. The bias at which $\nu_\mathrm{TG1}=0$ breaks down appears heavily dependent on the density under TG1 while the bias at which $\nu_\mathrm{TG1}=1$ breaks down is relatively independent of the density under TG1, occurring at $\pm V_\mathrm{EZ}$ across the plateau. \textbf{(D)} The dependence of d$I$/d$V$ on $V_\mathrm{dc}$ shows a sharp increase at $\pm V_\mathrm{EZ}$. \textbf{(E)}  The dependence of d$I$/d$V$ on $\nu_\mathrm{TG1}$ at $V_\mathrm{dc} = 0$ shows well quantized quantum Hall plateaus at $\nu_\mathrm{TG1}$ = 0, 1, and 2.}
		\label{fig:Supp2}
	\end{figure*}

	\begin{figure*}[t]
		\includegraphics[width=0.8\textwidth]{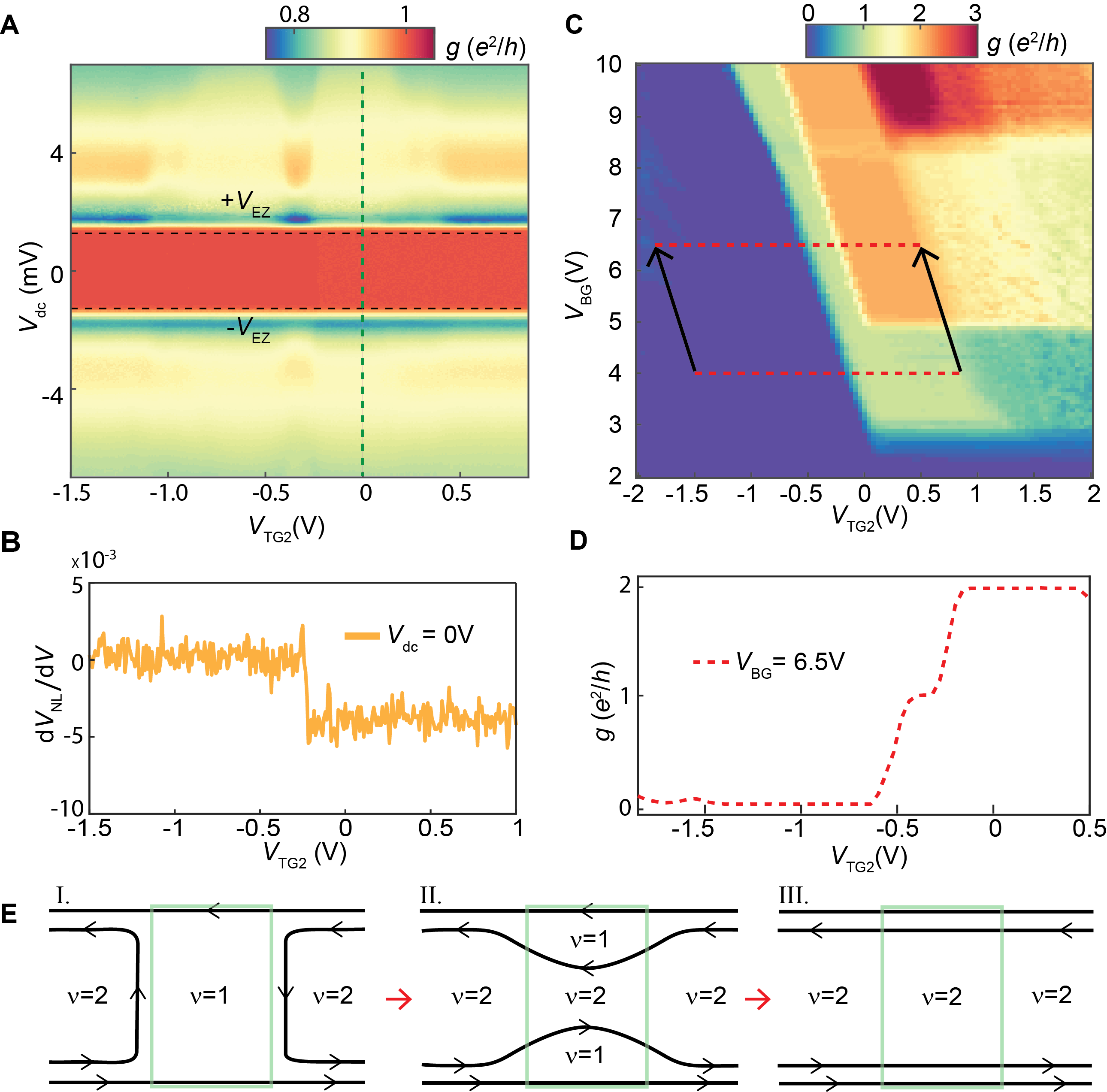}
			\caption{\textbf{The conditions under which non-local voltage ($S_\mathrm{NL}$) is measured.} \textbf{(A)} The conductance between L$_3$ and L$_2$ as a function of $V_\mathrm{dc}$ and $V_\mathrm{TG2}$ ($B$ = 8 T). Horizontal dashed black lines indicate $\pm V_\mathrm{EZ}$. Vertical green dashed line is where the line cut in Fig.\ 3B is taken. We see a sharp drop in conductance when $|V_\mathrm{dc}|> V_\mathrm{EZ}$ due to magnon generation. This drop is largely unaffected when top gate 2 (TG2) is changed.  Features at $|V_\mathrm{dc}|> V_\mathrm{EZ}$ coinciding with $\nu_\mathrm{TG2} = -1$ and $\nu_\mathrm{TG2} = 1$  indicate that magnons absorbed at the non-local voltage contacts affect the amount of magnons absorbed at the drain contact.  \textbf{(B)} $S_\mathrm{NL}$ is measured between L$_4$ and L$_5$ at $V_\mathrm{dc}$ = 0, showing a small negative voltage when the top gate is tuned from $\nu = 0$ to $\nu = 1$. This indicates a small number of bulk carriers that give a resistance between the two contacts --- a quantity which gives a small background to the $S_\mathrm{NL}$ signal, which can be subtracted out when calculating the value of $S_\mathrm{NL}$ when $|V_\mathrm{dc}|> V_\mathrm{EZ}$. \textbf{(C)} Two-terminal conductance measured between L$_3$ and L$_2$ as a function of the gate voltage on TG2 ($V_\mathrm{TG2}$) and on the back gate ($V_\mathrm{BG}$) ($V_\mathrm{dc}$ = 0). The line cut in Fig.\ 3C (main text) is meant to show the corresponding filling factors under TG2 for the voltage range on the x-axis, with $V_\mathrm{BG}$ = 4V (bulk at $\nu$ = 1). However, a line cut at $V_\mathrm{BG}$ = 4V does not show the transition between $\nu = 1$ and $\nu = 2$ because there is no equilibration between the $\nu = 1$ and $\nu = 2$ edges due to opposite spin polarization \cite{Amet2014, Wei2017}. We therefore use a line cut taken at $V_\mathrm{BG}$ = 6.5V ($\nu_\mathrm{BG} = 2$), where the step between $\nu = 1$ and $\nu = 2$ is clear, in order to estimate the steps in filling factor at $V_\mathrm{BG}$ = 4V. In order to account for the extra contribution in density due to the additional 2.5V applied by the back gate, we take the voltage interval of $V_\mathrm{TG2}$ at $V_\mathrm{BG}$ = 4V and shift it up by the slope of the hall plateaus (indicated by the black arrows pointing from the red-dashed line at 4V to the red-dashed line at 6.5V). \textbf{(D)} Conductance over the voltage range of $V_\mathrm{TG2}$ indicated by the red-dashed line in (C) at fixed $V_\mathrm{BG}$ = 6.5V.}
		\label{fig:Supp7A}
	\end{figure*}

\begin{figure*} [t]
	\ContinuedFloat
	\includegraphics[width=0.9\textwidth]{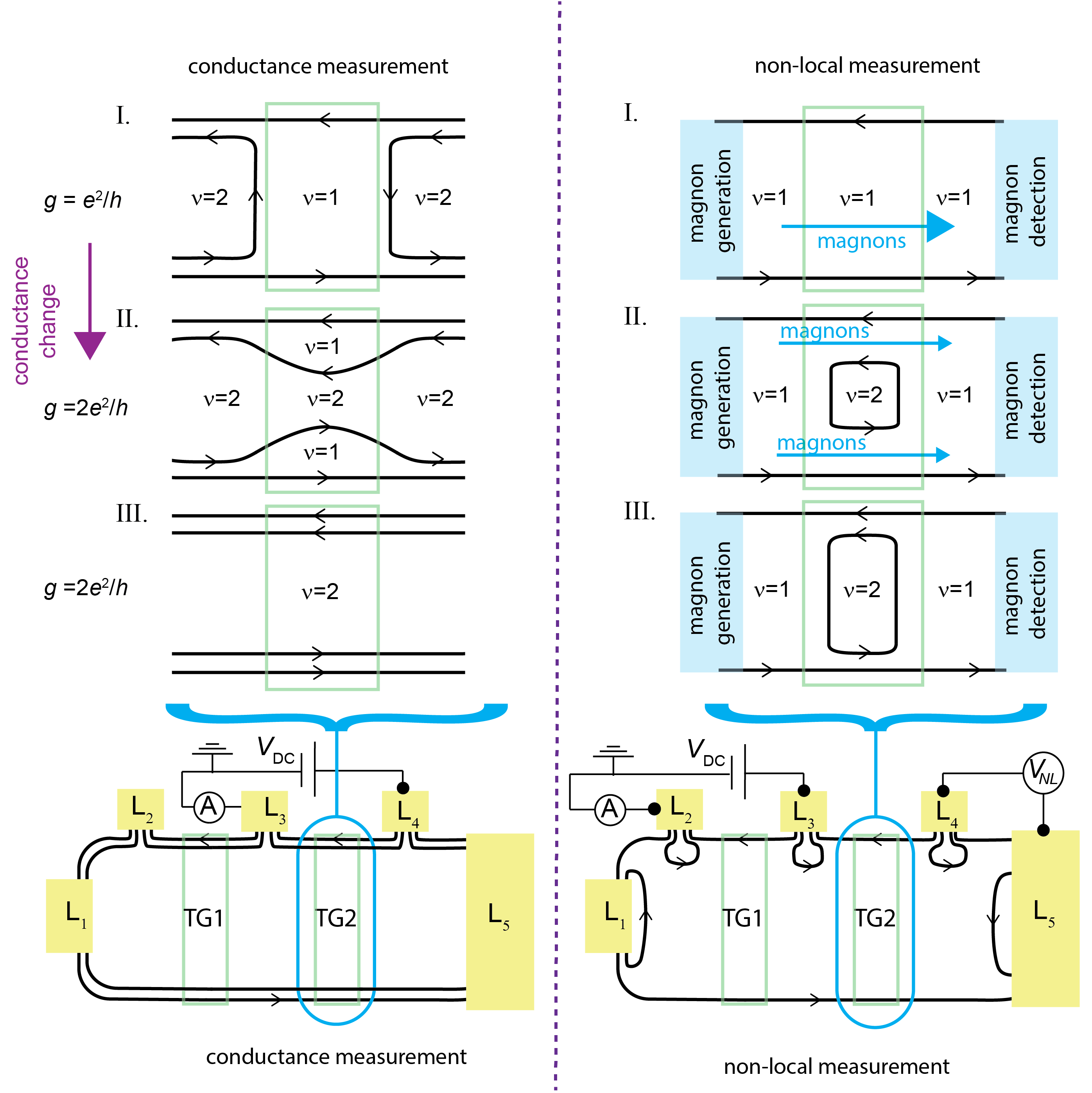}
	\caption{\textbf{(E) Comparison of a two-terminal conductance measurement across a top gate (left) to a non-local magnon-transmission measurement across the same top gate(right), as the density in the top-gated region is tuned from $\nu = 1$ to $\nu = 2$.} Panel I: (left) the two edge states in the outer regions are not yet able to enter the top-gated $nu$ = 1
	region, resulting in a conductance of $e^2/h$. In the corresponding non-local measurement
	(right), magnons are able to propagate, yielding a non-local voltage. Panel II: (left) As the density is increased further, the developing $\nu = 2$ region under the top gate connects with the outer $\nu = 2$ regions, changing the measured conductance to $2e^2/h$. However, some $\nu = 1$ regions under the top gate remain present. In the corresponding non-local measurement (right), these remaining  $\nu = 1$ regions under the top gate still allow magnon transport. In the non-local measurements shown in Fig. 3C, we expect these regions to be responsible for the non-local voltage signal seen when the region under the top gate is transitioning from $\nu_{TG2}$ = 1 to $\nu_{TG2}$ = 2. Panel III: Once the density is increased sufficiently, the topgated region consists almost entirely of $\nu = 2$, yielding a conductance of $2e^2/h$ and a near complete suppression of magnon transport in the non-local measurement.}
	\label{fig:Fig7E}
\end{figure*}

	\begin{figure*}[t]
		\includegraphics[width=0.8\textwidth]{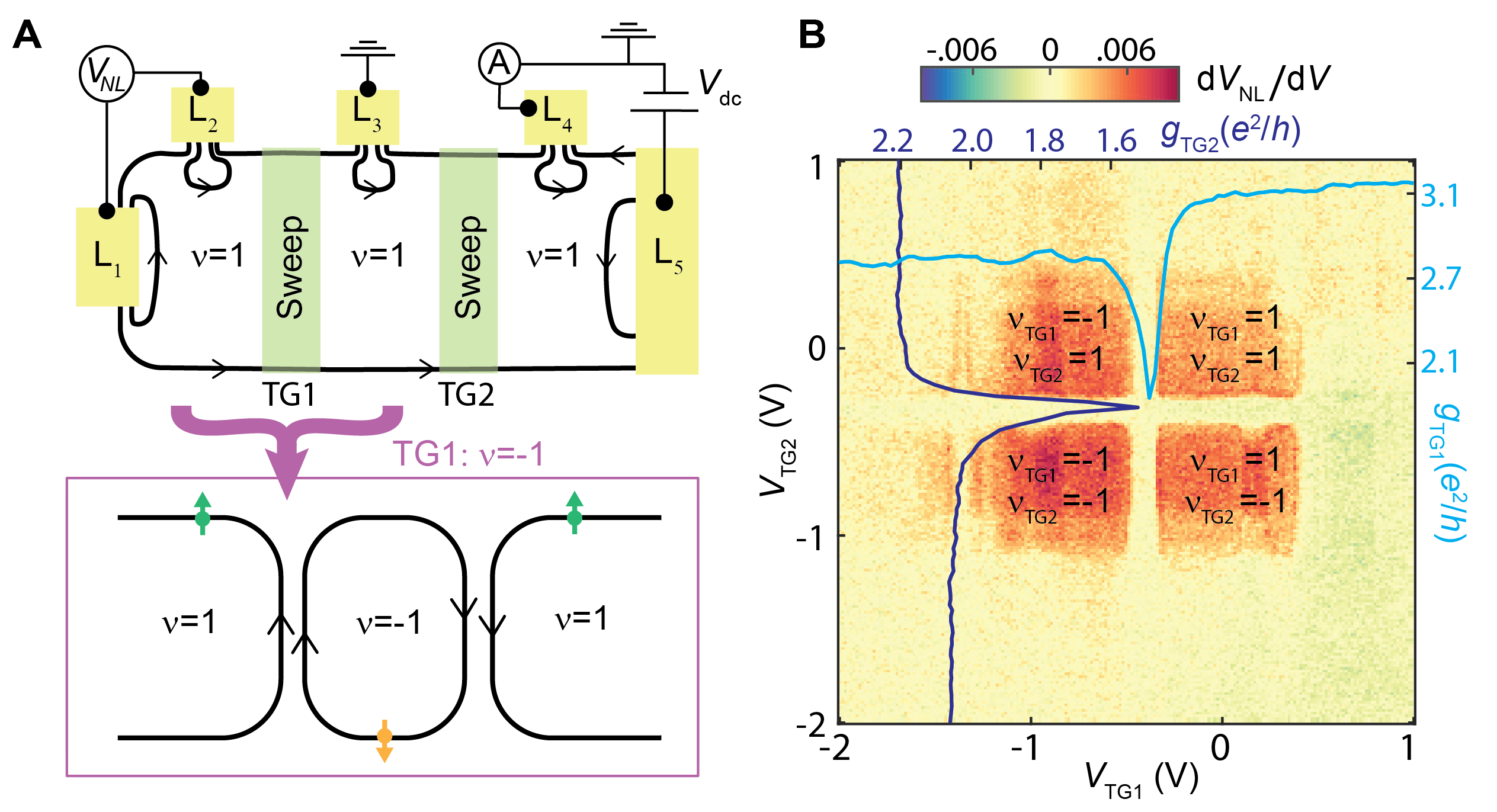}
		\caption{\textbf{Dependence of $S_\mathrm{NL}$ on filling factors under TG1 and TG2.} \textbf{(A)} A circuit configuration for measuring a non-local voltage in device 2 (schematic). The filling factor under TG1 ($\nu_\mathrm{TG1}$) and under TG2 ($\nu_\mathrm{TG2}$) are both swept from -2 to 2, while the outside regions are maintained by a fixed back-gate voltage at $\nu_\mathrm{BG}$ =1 ($V_BG$ = 4V, B = 8 T). The bottom panel highlights the case of $\nu_\mathrm{TG1}$=-1: Edge states in both regions co-propagate along the boundary, but do not equilibrate because of their opposite spin-polarization \cite{Wei2017}. \textbf{(B)} Setting $\mu > E_\mathrm{Z}$ ($V_\mathrm{DC}$ = -2.8 mV), and measuring $S_\mathrm{NL}$ between L$_2$ and L$_1$ we find strong non-local signals in four quadrants around $\nu = 0$. Strips where the signal is highly suppressed coincide with where the charge neutrality point occurs in density measurements of TG1 (TG2) at B = 0 T, shown by the superimposed light blue (dark blue) line cuts. We see similar signals in all four quadrants, indicating that magnons are not suppressed by the $\nu = -1$ regions.}
		\label{fig:Supp2}
	\end{figure*}

\clearpage

	\begin{figure*}[t]
		\includegraphics[width=1.0\textwidth]{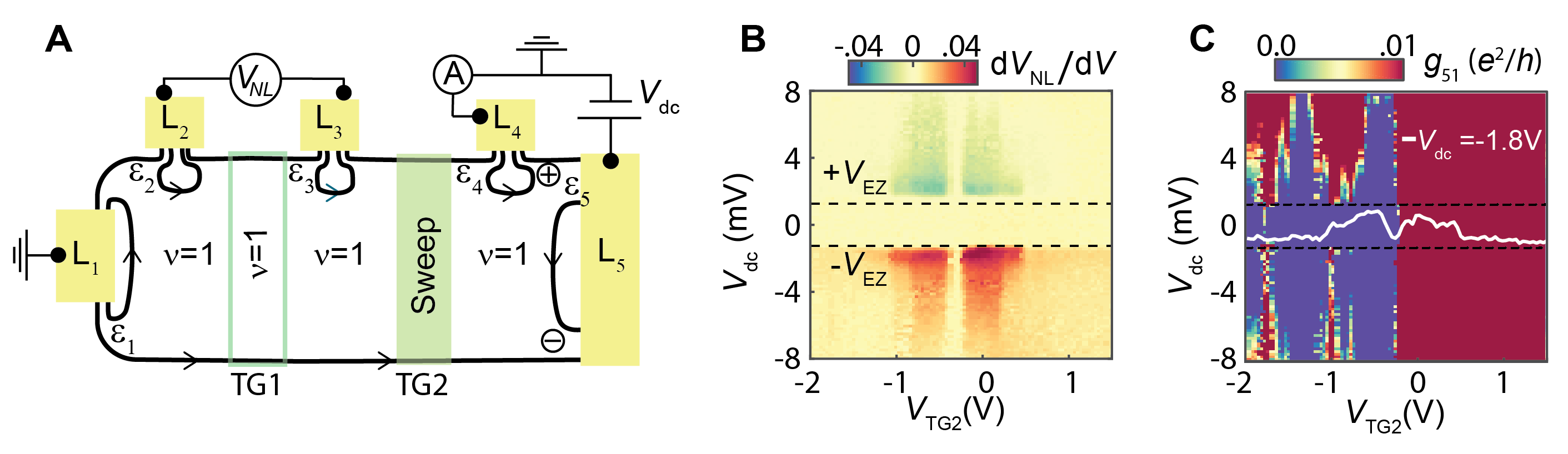}
		\caption{\textbf{Absence of current leakage when spin transport is mediated by the $\nu = -1$ ferromagnet.} \textbf{(A)} A circuit used to measure $S_\mathrm{NL}$ in device 2, as well as a leakage current across TG2 when it is tuned to $\nu_\mathrm{TG2} = -1$ (schematic). $\nu_\mathrm{TG1} = 1$ for all measurements while $\nu_\mathrm{TG2}$ is swept from -2 to 2. \textbf{(B)} Non-local voltage ($S_\mathrm{NL}$) measured between L$_3$ and L$_2$ as a function of $V_\mathrm{dc}$ and $V_\mathrm{TG1}$. Horizontal dashed black lines indicate $\pm V_\mathrm{EZ}$. We note a delay in the onset of the non-local signal for positive bias, which we tentatively attribute to the fact that the absorption of magnon generation for positive bias is far from both the non-local leads and is mostly absorbed at $\varepsilon_5$ and $\varepsilon_4$, with only enough magnons to generate a non-local signal at larger energies.  \textbf{(C)} Conductance into L$_1$ with the color scale saturated. This measures the current not drained at L$_4$ due to the contact resistance $R_\mathrm{C}$. Black dashed lines indicate  $\pm V_\mathrm{EZ}$. White line cut is taken from the plot shown in (B) (over the same span of $V_\mathrm{TG1}$) at fixed $V_\mathrm{dc}$ = -1.8mV, and overlaid onto the conductance map. When we see an increase in $S_\mathrm{NL}$ there is a negligible amount of leakage current ($g_{51} < 0.01$ $e^2/h$) measured at L$_1$. Additionally, when we see an increase in the leakage current ($g_{51} > 0.01$ $e^2/h$), there is no corresponding effect on $S_\mathrm{NL}$. This is expected because an edge current should bring L$_3$ and L$_2$ to the same chemical potential. From this we conclude that the $S_\mathrm{NL}$ we measure is not due to leakage current.}
		\label{fig:Supp2}
	\end{figure*}

	\begin{figure*}[t]
		\includegraphics[width=1.0\textwidth]{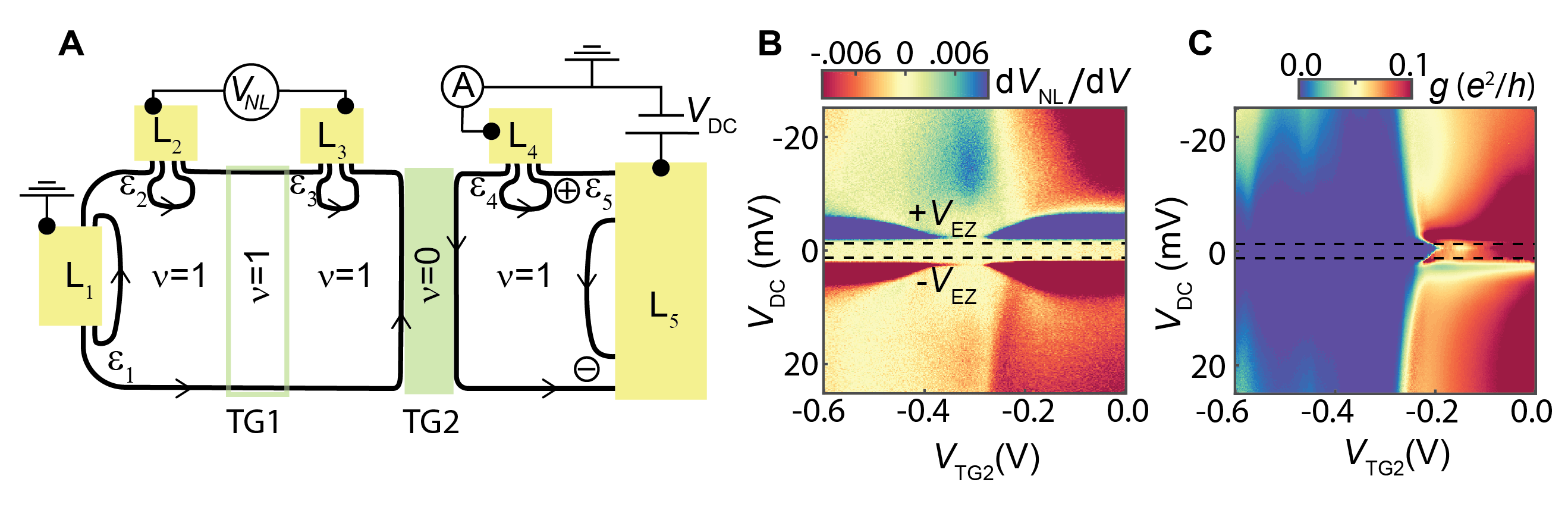}
		\caption{\textbf{Absence of current leakage when spin transport is mediated by the $\nu = 0$ CAF.} \textbf{(A)} The circuit used to measure $S_\mathrm{NL}$ in device 2 across a $\nu = 0$ region (schematic). L$_1$ is grounded in order to measure the amount of residual charge that leaks through to the other side of $\nu_\mathrm{TG2} = 0$. \textbf{(B)}  When magnons are generated in the $\nu_\mathrm{BG}$, $\nu_\mathrm{TG1}$ = 1 region and $\nu_\mathrm{TG2}$ = 0, we see an onset of $S_\mathrm{NL}$ at energies exceeding $\pm V_\mathrm{EZ}$. This indicates that higher energy magnons have overcome the interface barriers and have propagated through the $\nu_\mathrm{TG2}$ = 0 region. At more positive gate voltages, we see the effects of the residual current on $S_\mathrm{NL}$ (due to the finite contact resistance of L$_4$) which passes through when $\nu_\mathrm{TG2} > 0$. We observe that these effects disappear once $\nu_\mathrm{TG2} > 0$ and do not play a role in the $S_\mathrm{NL}$ signal measured in this region.  \textbf{(C)} Residual current measured at L$_1$ indicating that residual leakage does not correlate with the appearance of the $\nu_\mathrm{TG2} = 0$ signal.}
		\label{fig:Supp2}
	\end{figure*}

	\begin{figure*}[t]
		\includegraphics[width=1.0\textwidth]{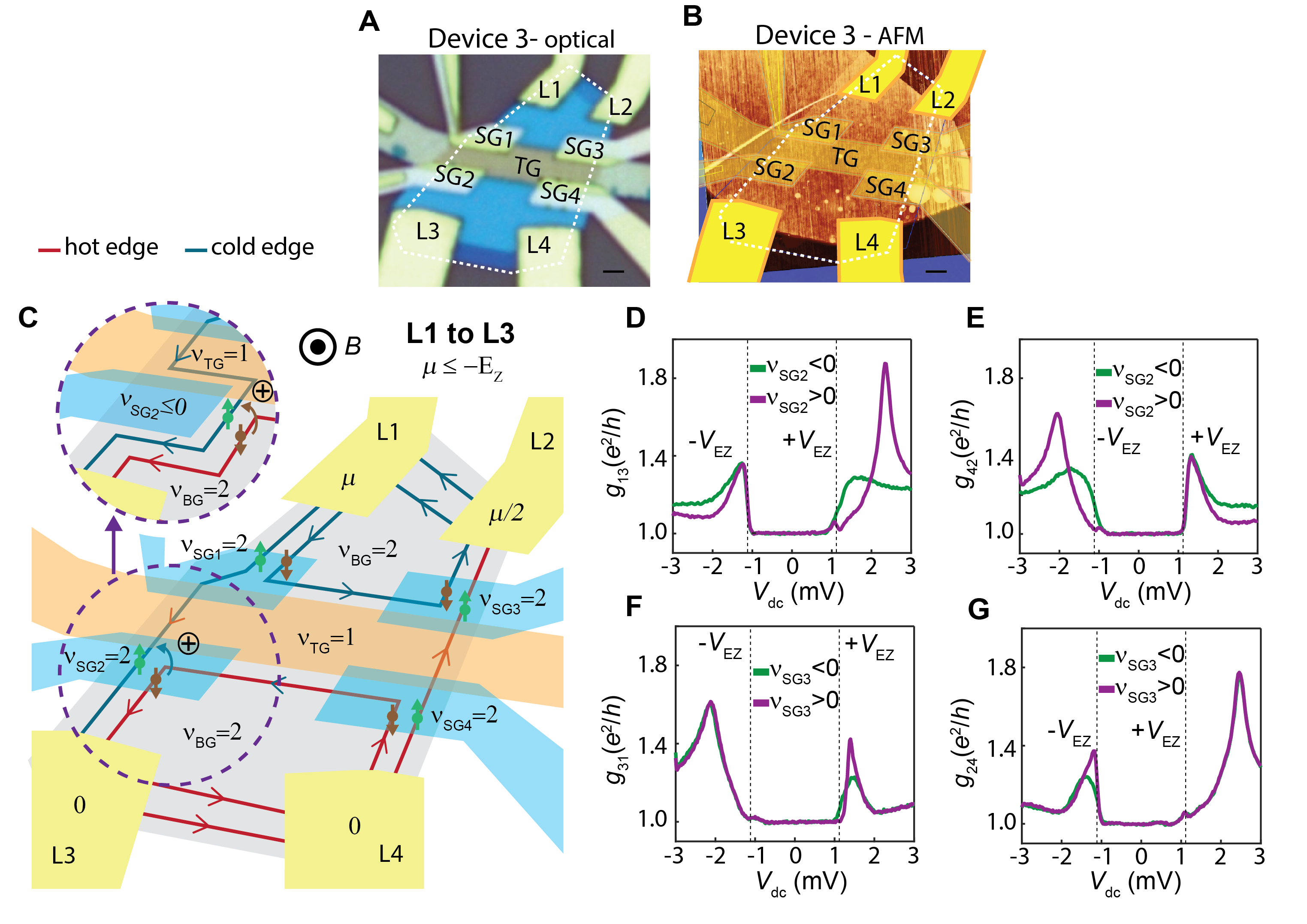}
		\caption{\textbf{Verifying positive and negative bias magnon generation locations (device 3).} \textbf{(A)} Optical micrograph of device 3. The outline of the graphene is shown by the dotted white line, and the scale bar is 1 $\mu$m. For this device, an extra BN dielectric (10nm) was used between the top gates and side gates to electrically isolate them. There are 4 leads (L$_1$-L$_4$), one top gate (TG), and four side gates (SG1 - SG4). \textbf{(B)} AFM image of device 3. \textbf{(C)} Schematic of device 3 depicting a two-terminal conductance measurement between L$_1$ (source) and L$_3$ (drain) with L$_2$ and L$_4$ floating. The leads are yellow, the top gate (TG) is orange and the side gates are light blue. The regions outside of the top-gated region (including the side gates) are tuned to $\nu = 2$ and the region under the top gated region is tuned to $\nu = 1$. Chiral edge states are shown by the lines with arrows and edges with higher (lower) chemical potential are colored red (blue) and labeled hot (cold). The side gates can be used to push the edge states away from the physical edge of the device (as illustrated in the inset, for SG2). \textbf{(D)} Two-terminal conductance measurement at $B$ = 7 T where a constant d.c. voltage ($V_\mathrm{dc}$) and a 50 $\mu \mathrm{V}$ a.c. excitation voltage are applied to L$_1$ (source) and the differential conductance (d$I$/d$V$) is measured through L$_3$ (drain). Magnons are generated when a spin-down hot edge meets a spin-up cold edge at $E_\mathrm{Z}$. For this configuration, we expect magnons to be generated under SG2 for positive $V_\mathrm{dc}$ only. When we reach $+V_\mathrm{EZ}$ we see a change in the conductance while at $-V_\mathrm{EZ}$ we see almost no change, as expected. \textbf{(E-G)} Similar analysis for different lead configurations shows magnons are generated in accordance with our model predictions. The effect of SG2 is stronger than SG3 for unknown reasons. The exact change in conductance is difficult to predict because we are changing both the nature of scattering between the two edge states \cite{Wei2017} as well as the distance between magnon generation and absorption, so here we note only qualitative changes.}
		\label{fig:Supp2}
	\end{figure*}
		
	\end{document}